\documentclass[twocolumn]{aastex61}
\usepackage[utf8]{inputenc}
\usepackage{graphicx}
\usepackage{amsmath}
\usepackage{epstopdf}
\usepackage{mathrsfs}
\usepackage{natbib}

\begin{document}
\title{Two procedures to flag radio frequency interference in the UV plane}

\author{Srikrishna Sekhar}
\email{sekhars@students.iiserpune.ac.in}
\affil{Indian Institute of Science, Education and Research, Pune}

\author{Ramana Athreya}
\correspondingauthor{Ramana Athreya}
\email{rathreya@iiserpune.ac.in}
\affil{Indian Institute of Science, Education and Research, Pune}

\begin{abstract}

   We present two algorithms to identify and flag radio frequency interference
   (RFI) in radio interferometric imaging data. The first algorithm utilizes the
   redundancy of visibilities inside a UV cell in the visibility plane to identify
   corrupted data, while varying the detection threshold in accordance with the
   observed reduction in noise with radial UV distance.  In the second algorithm,
   we propose a scheme to detect faint RFI in the visibility time-channel plane of
   baselines. The efficacy of identifying RFI in the residual visibilities is
   reduced by the presence of ripples due to inaccurate subtraction of the
   strongest sources. This can be due to several reasons including primary beam
   asymmetries and other direction dependent calibration errors. We eliminated
   these ripples by clipping the corresponding peaks in the associated Fourier
   plane. RFI was detected in the ripple-free time-channel plane but was flagged
   in the original visibilities. Application of these two algorithms to 5
   different 150 MHz datasets from the GMRT resulted in a reduction in image noise
   of 20-50\% throughout the field along with a reduction in systematics and a
   corresponding increase in the number of detected sources.  However, on
   comparing the mean flux densities before and after flagging RFI we find a
   differential change with the fainter sources ($25\sigma <$ S $< 100$ mJy)
   showing a change of -6\% to +1\% relative to the stronger sources (S $>$ 100
   mJy).  We are unable to explain this effect but it could be related to the
   CLEAN bias known for interferometers.

\end{abstract}


\section{Introduction} \label{sec:intro}

At low radio frequencies, below 1 GHz, the radio frequency interference
(RFI) environment is quite active and can cause severe degradation in image
quality. RFI can increase image noise by up to an order of magnitude above
the thermal noise expected for the telescope. The images also usually
exhibit widespread systematics in the form of multiple ripples which increase
the detection thresholds of faint objects.

There has been much effort directed towards RFI mitigation and flagging
strategies in both hardware (pre-correlation) and software (post-correlation)
regimes.  Hardware based techniques typically use either some sort of
a reference signal to measure the interference and subtract it from the data
\citep{briggs_removing_2000, barnbaum_new_1998,
hellbourg_reference_2014,fridman2001rfi}, {or use specialised hardware
such as additional antennas (or antenna arrays) to null sources of
interference along certain directions \citep{van2004spatial,kocz2010radio}.
Among the software based tools there are techniques that attempt to excise
the RFI from the data \emph{i.e.,} recover the uncorrupted visibilities
\citep{athreya_new_2009, golap_post_2005, pen_gmrt_2009, offringa2012post} and
methods to remove - \emph{i.e.}, flag - the affected data
\citep{offringa_morphological_2012, offringa_post-correlation_2010,
bhat2005radio, middelberg2006automated, winkel2007rfi}. The software methods
have an advantage in that they can be applied to both new as well as archival
observations.

In general a single mitigation strategy has not proved to be very effective
since different sources of RFI leave different signatures in the data.
Persistent RFI appears as strong fringes in a baseline, and the amplitude and
phase of these fringes can change as a function of both time and frequency.
Broadband RFI can affect the entire baseline and cause fluctuations in antenna
gain and are much more difficult to characterise and eliminate. Intermittent
RFI, localized in time and frequency, look like `hotspots' in the visibility
data, and cause large scale ripples in the image plane.

We describe here two new methods to identify and flag intermittent RFI,
which have consistently yielded high sensitivity images when applied to
a variety of GMRT observations.

\section{Description of the algorithms}

The measured visibilities in any polarization in the presence of
RFI signal can be written as:

\begin{align}
   V^o = \text{G}_i   \text{G}_j^* (V^\text{sky}_{ij}
   + V^\text{RFI}_{ij}e^{if_{ij}})
   + \eta_{ij}
\end{align}

\noindent
where $ij$ are antenna indices, $\text{V}_{ij}^\text{sky}$ are the visibilities
due to sky emission, $\text{V}_{ij}^{RFI}$ is correlated RFI, and
$e^{if_{ij}}$ is the fringe stopping function (and the only
polarization independent factor), $\eta_{ij}$ is additive noise in the system.
$\text{G}_i$ and $\text{G}_j$ are the complex antenna gains, which can be
affected by strong RFI  even if the RFI is uncorrelated.  $e^{if_{ij}}$ will
stop the fringe of the cosmic source at the phase centre, but introduces
a corresponding fringe on a stationary (terrestrial) source, like RFI.
\citet{athreya_new_2009} used the form of $e^{if_{ij}}$ to excise RFI while
recovering the visibilities.

$V^\text{RFI}$ can be orders of magnitude larger than $V^\text{sky}$, and vary
with time, frequency and baseline. The second term in the above
equation causes poorer solutions during self-calibration and introduce
systematic errors in the image.

In this paper we focus on intermittent RFI which are localized in the
visibility space.  These localized hotspots will result in large scale ripples
across the image. We explore two different visibility spaces,
\emph{viz.} the binned UV plane of the entire interferometer and the
time-channel plane of a single baseline, to locate and flag corrupted
visibilities. The two algorithms are individually called GRIDflag and TCflag,
respectively, and are combined into an integrated RFI flagging package called
IPFLAG.

\subsection{RFI flagging in the gridded UV plane - GRIDflag}

Visibilities sampled by a baseline form a continuous track across the UV plane.
The tracks of different baselines are distributed irregularly and usually
sparsely. Imaging algorithms compensate for this by interpolating these sampled
visibilities onto a regular grid to be able to use the fast Fourier transform
(FFT) algorithm \citep{thompson2001interferometry}. This gridded UV plane is
fundamental to almost all imaging algorithms, and the sampled cells within the
UV grid define the UV coverage of the observation. The size of the UV-cell is
related to the field of view being imaged.  A single baseline will in general
contribute multiple visibilities to a particular cell.  These multiple
visibilities usually lie within a short time interval of each other, unless the
observation spans multiple epochs.  Multiple baselines may contribute to the
same cell, but at different times.

\begin{figure}[htbp]
   \fig{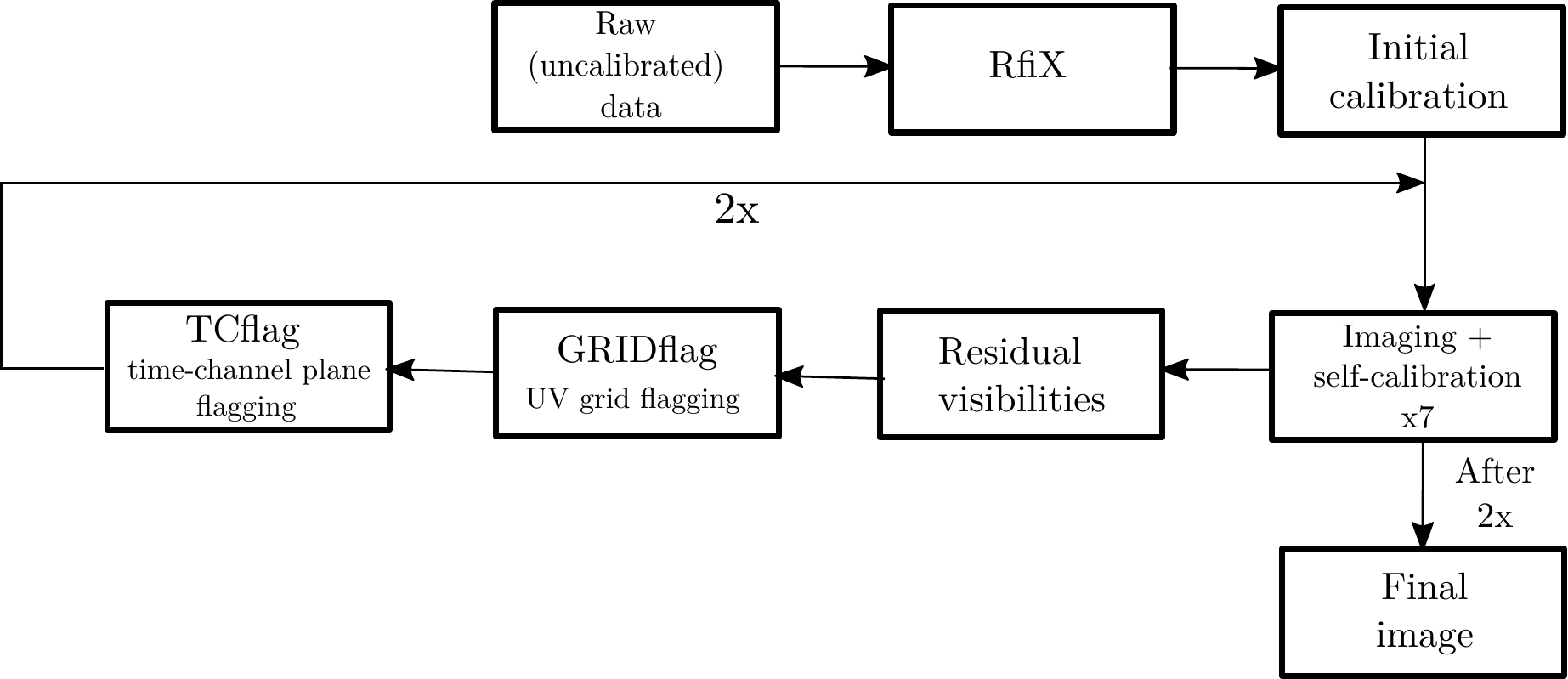}{\linewidth}{}
   \caption{Flowchart describing the data
   analysis recipe which was used to obtain the final data sets shown in the
   comparisons below.}
   \label{fig:flowchart}
\end{figure}

All the visibility samples within a cell approximately measure the same
celestial information, but differ in the RFI environment that they encountered
due to different times of observation.  We propose to use this dichotomy to
identify and flag RFI affected visibilities.

The first step of the GRIDflag algorithm is to bin the visibilities
based on their UV coordinates. The size of a UV-bin is similar in size to the
UV-cell used while imaging. We assume that in the absence of RFI the
differences between the visibilities in a UV-bin is dominated by system
temperature and not source structure. This is valid when applying the
scheme to the residual visibility plane obtained after subtracting the strongest
sources. Thus, any differences between the visibilities within a UV-bin
component which are well in excess of the system noise can be ascribed to RFI.
The visibility function must be locally smooth for any realistic sky intensity
distribution.  Therefore, one can combine data from adjacent UV-bins to
calculate statistically secure thresholds to identify RFI.

The standard radio astronomy imaging procedure consists of pre-calibration
flagging, and several rounds of imaging and self-calibration followed by (often
manual) residual visibility flagging procedures available in CASA/AIPS
\citep{mcmullin2007casa, greisen2003aips}.  Typically, observers using the GMRT
150 MHz band produce images with RMS noise of 1.5 - 5 mJy/beam by using these
standard procedures. We have routinely reached below 1 mJy/beam using the
pre-calibration RfiX procedure \citep{athreya_new_2009} to excise persistent
broad-band RFI. We applied the algorithms described in this paper at the end of
this standard procedure.  Our recipe is as follows:

\begin{enumerate}
   \item Apply the RfiX algorithm and the standard CASA/AIPS calibration, imaging
      and flagging process to obtain the residual visibilities.

   \item Allot the visibilities into bins in the UV plane. These bins are
      approximately the same size as the UV-cells used for gridding by imagers.
      Calculate the robust median and standard deviation (with respect to the
      median) of all the residual visibilities falling within each UV-bin.

   \item Partition the UV plane into several annuli and calculate RFI
      thresholds as a function of UV radius. This is because both RFI and source
      signal in the residual visibilities tend to decrease with radial distance.
      The choice of the annulus width is not critical and is decided by the
      competing requirements of tracking the change in RMS with radius and having
      sufficient UV-bins within an annulus.

   \item Use the distribution of medians in an annulus to exclude highly
      contaminated UV-bins while determining the smoothed median
      background surface. Note that this will be a function of UV radius.

   \item The smoothed median surface and the local standard deviation is
      used to identify RFI affected data within each UV-bin through any
      thresholding scheme - we used the visibility RMS to define the threshold.
      One can set this threshold either using data from within the same UV-bin
      or by combining other UV-bins in the immediate neighbourhood.

   \item Apply these flags to the original, un-smoothed and un-binned
      data, and redo the entire process of imaging and self-calibration.

\end{enumerate}

\begin{figure*}[htbp]
   \gridline{
      \fig{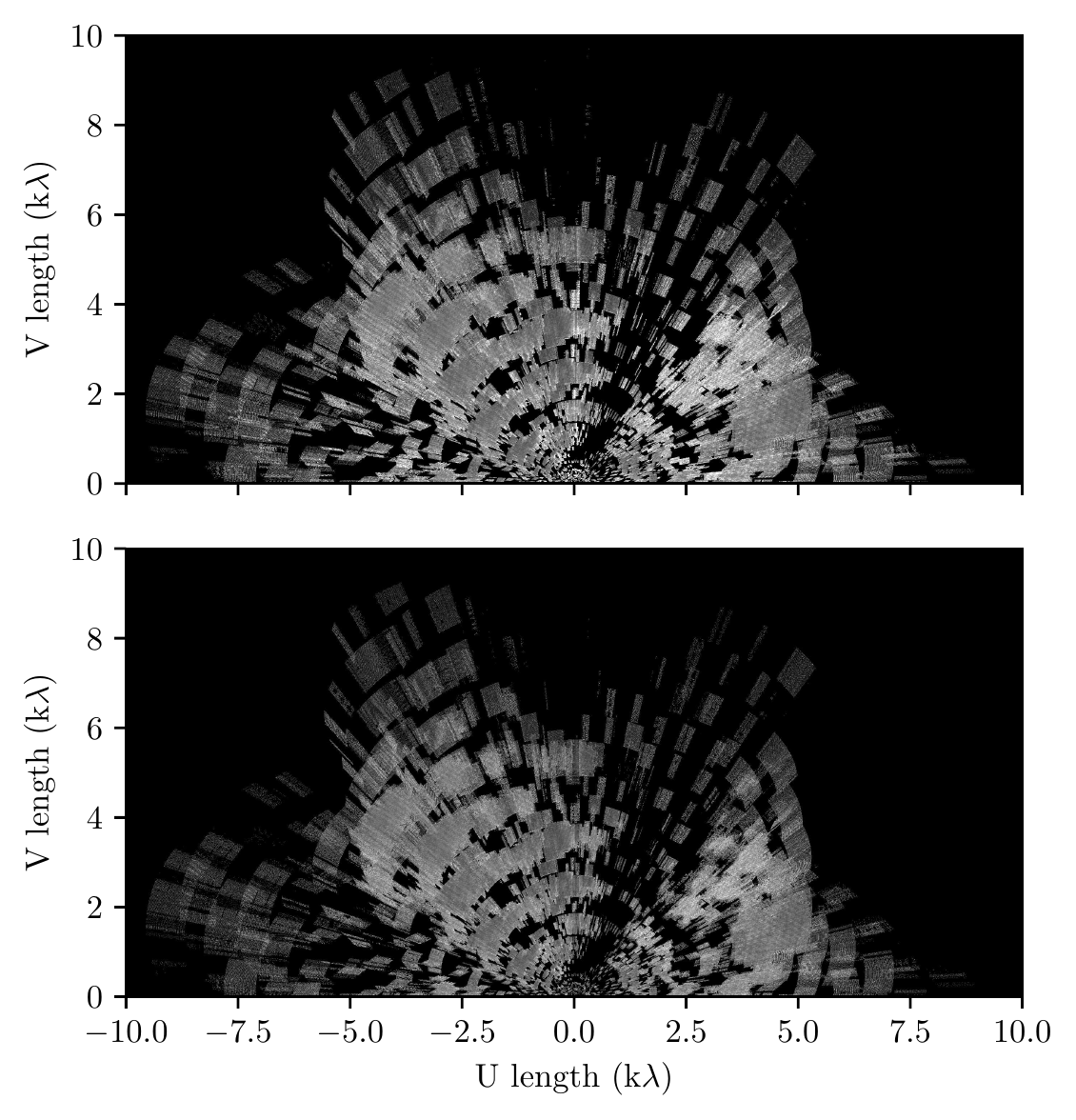}{0.325\linewidth}{3C286}
      \fig{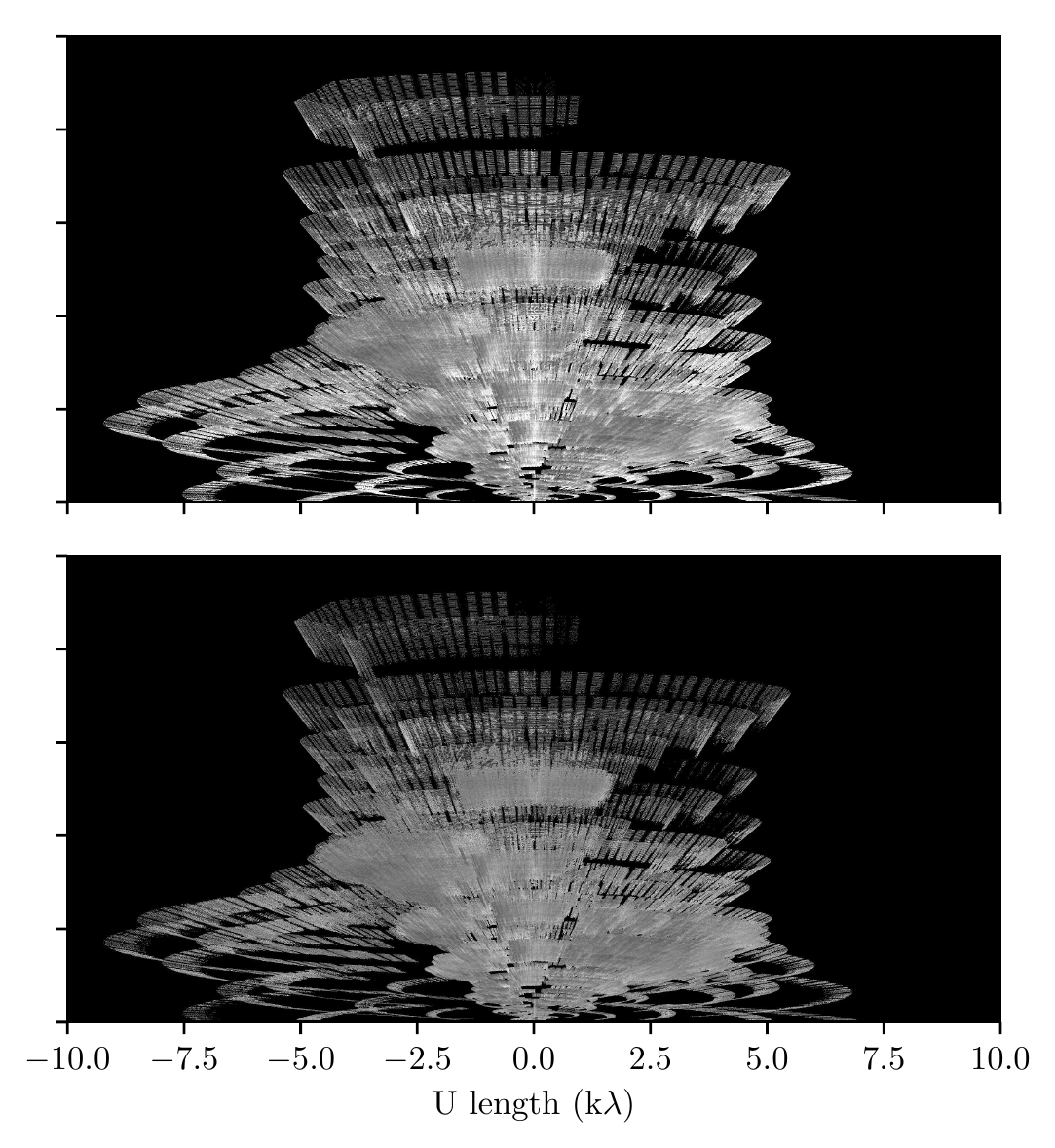}{0.31\linewidth}{VIRMOSC}
      \fig{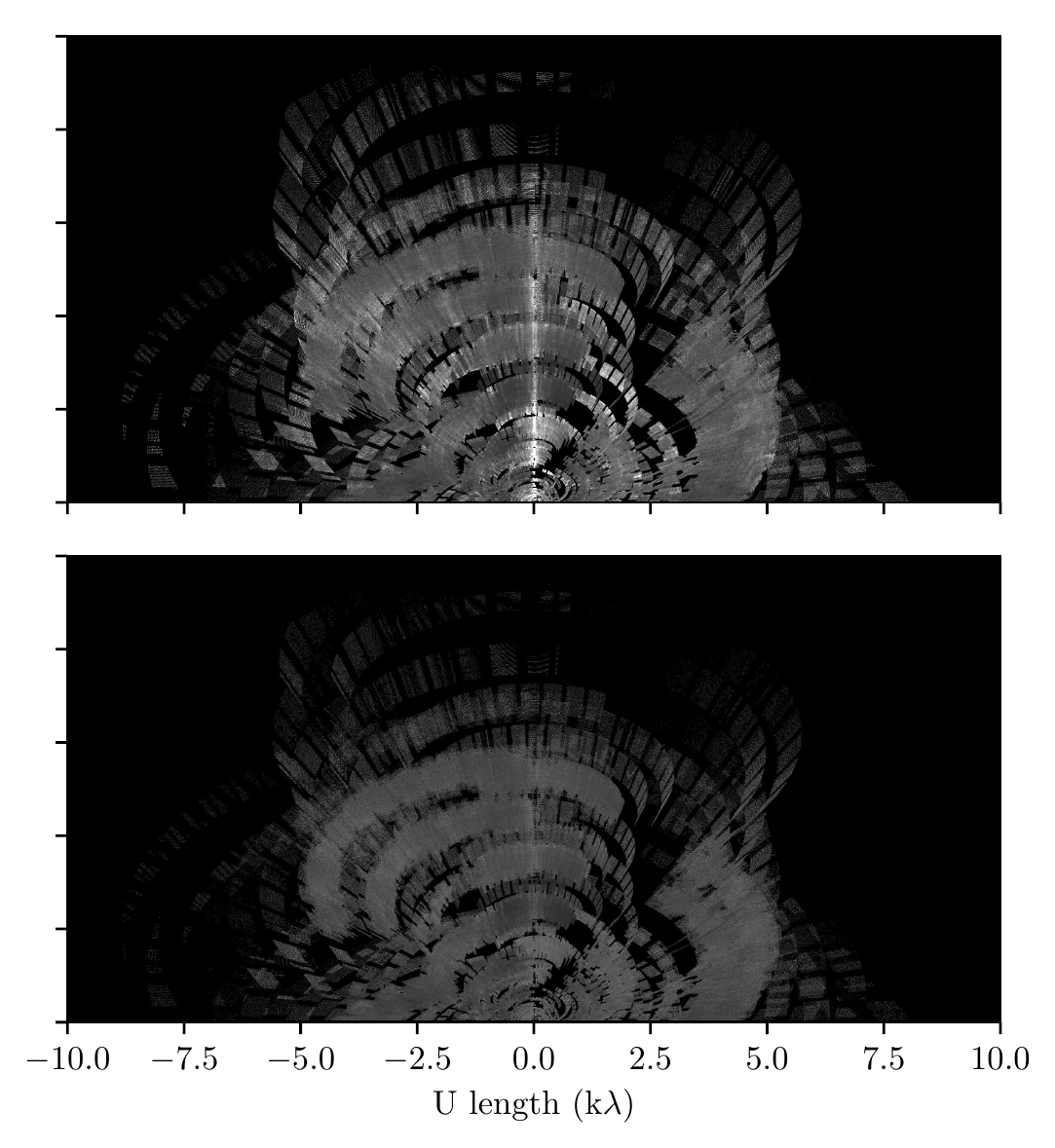}{0.31\linewidth}{J1453+3308}
   }
   \gridline{
      \fig{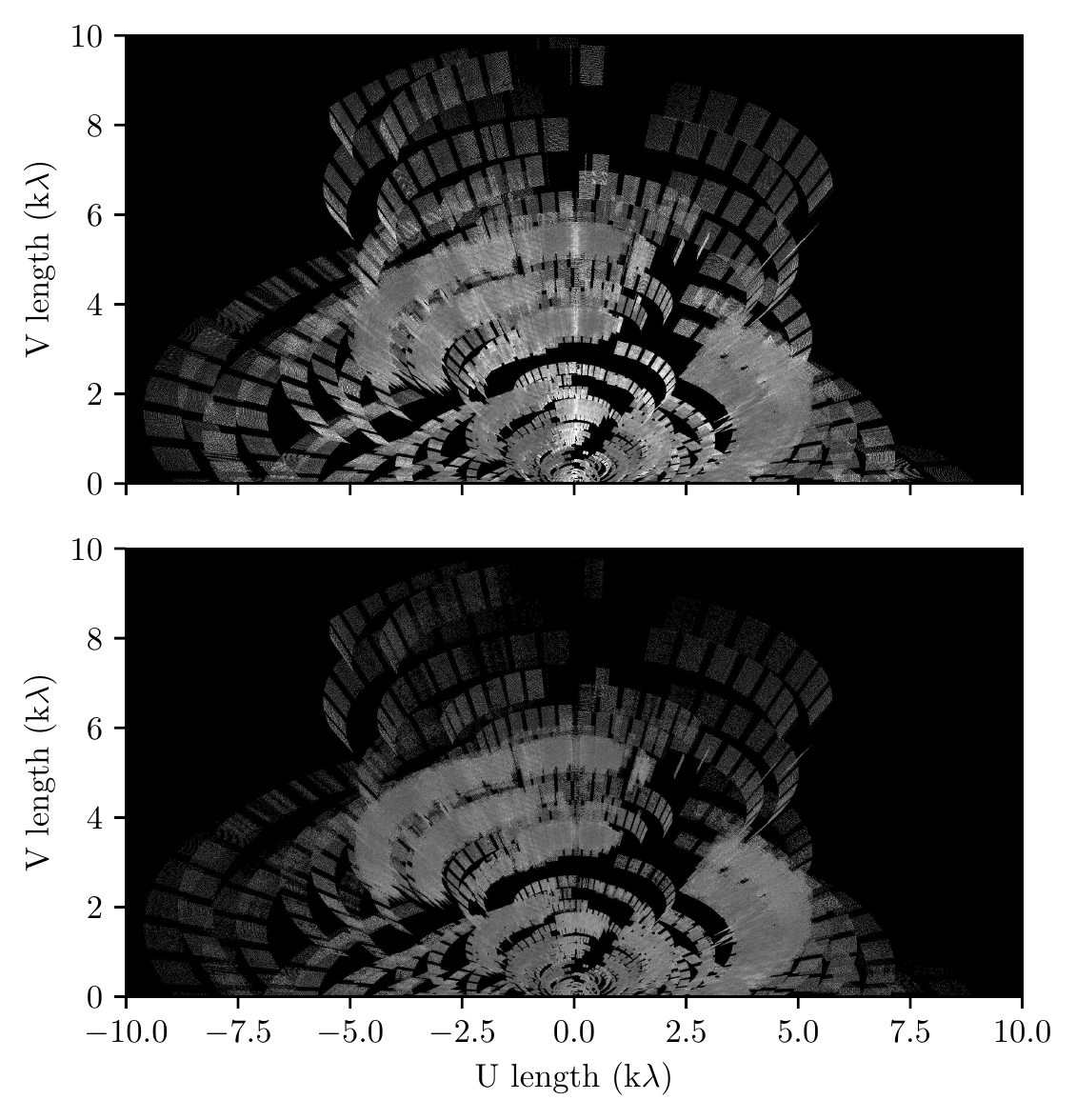}{0.325\linewidth}{J1158+2621}
      \fig{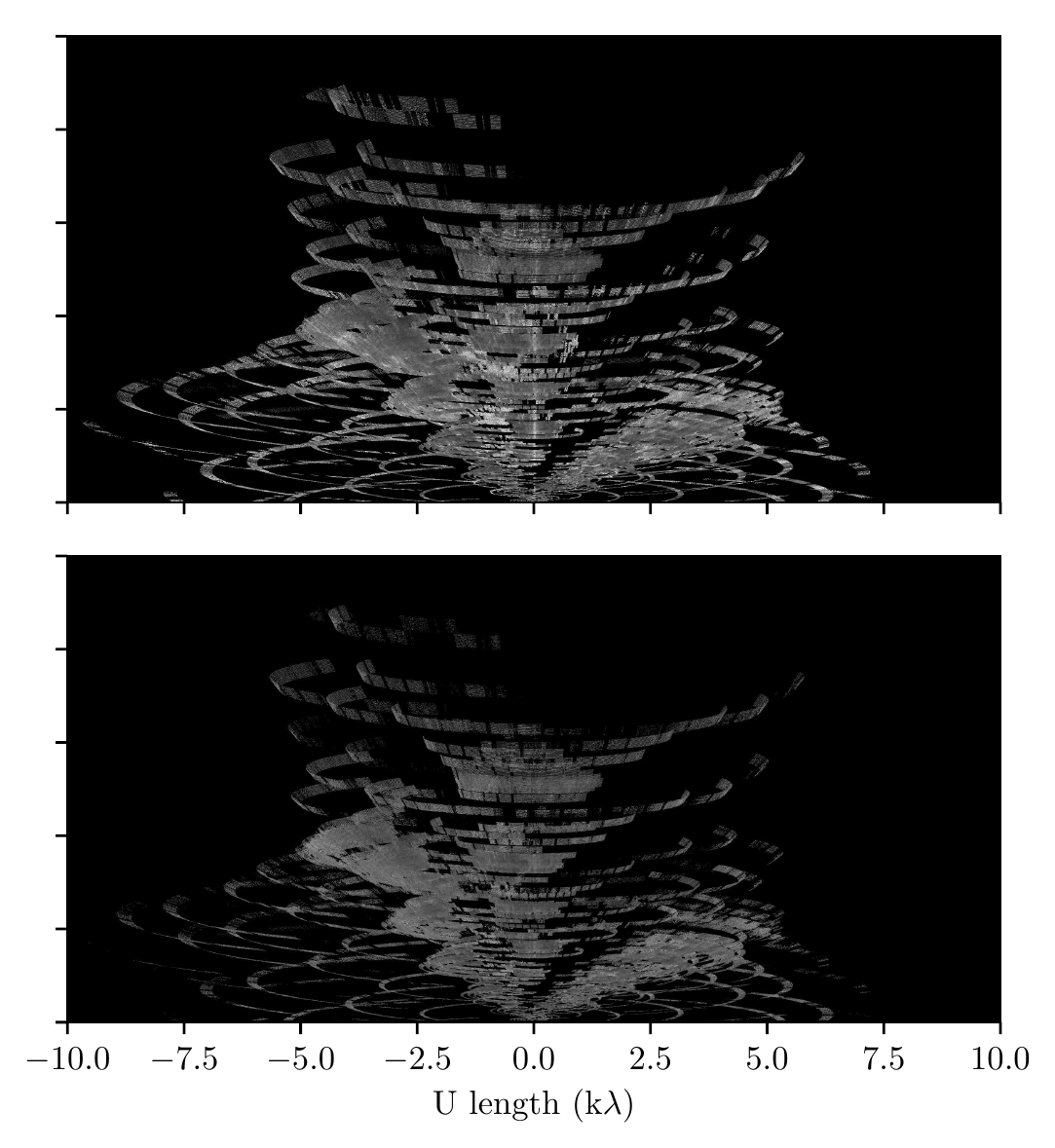}{0.31\linewidth}{A2163}
   }
   \caption{Comparison of the median binned UV plane before and
   after GRIDflag. The grayscale represents the
   median flux density in a $10\lambda \times 10\lambda$ UV-bin.}
   \label{fig:gridded_uv_plane}
\end{figure*}

This procedure largely preserves the UV coverage for two reasons: the flagging
in each UV-bin is processed separately and in most cases at least a few
visibilities in each UV-bin survive the process.  Secondly, this procedure
allows for smooth variation of standard deviation even within an annulus. Though
we have yet to implement it, the variation of standard deviation may be
compensated by differentially weighting the visibilities.

As a matter of detail, only one half of the visibility data is recorded since
the other half is simply a Hermitian conjugate.  Therefore the visibilities have
to be appropriately conjugated to ensure that they all lie within the same half
plane. One will also need to extend the data into a few UV-bins in the
other half for statistics at the edge.

We applied this procedure separately and successively for the amplitude, real,
and imaginary components in the RR and LL and stokes V polarisation modes.

\begin{deluxetable*}{ccccc|cccccc}
   \tablewidth{0.7\linewidth}
   \tablecolumns{5}
   \tablecaption{\label{table:image_parameters}The effect of IPFLAG on image
   parameters. The values of the dirty beam major and minor axes in arcsec,
   median image noise $\sigma_o$ and $\sigma_f$ in mJy/beam, and the number
   of detected sources $N_o$ and $N_f$ are listed for each field. We have also
   listed the percentage of UV-bins lost and visibility data flagged after
   IPFLAG.}
   \tablehead{
      \colhead{Source name} & \multicolumn{4}{c}{Without IPFLAG}
   & \multicolumn{6}{c}{With IPFLAG}\\
   \hline
   \colhead{} & \colhead{Major axis}
   & \colhead{Minor axis} & \colhead{$N_O$}
   & \colhead{$\sigma_O$} & \colhead{Major axis}
   & \colhead{Minor axis} & \colhead{$N_f$}
   & \colhead{$\sigma_f$} & \colhead{Flagged bins \%} & \colhead{Flagged points
   \%}
   }

   \decimals
   \startdata
   3C286      & 18.4 & 13.1 & 374 & 1.12 & 18.9 & 13.0 & 403 & 1.01 & 1.2
   & 2.4\\
   VIRMOSC    & 23.9 & 15.9 & 846 & 0.55 & 25.6 & 15.8 & 949 & 0.42 &  1.7
   & 6.4\\
   J1453+3308 & 20.2 & 14.5 & 862 & 0.46 & 18.9 & 14.1 & 870 & 0.38 & 2.7
   & 15.7\\
   J1158+2621 & 19.0 & 15.1 & 819 & 0.72 & 18.9 & 14.6 & 844 & 0.55 & 3.9
   & 15.5\\
   A2163      & 28.9 & 14.3 & 419 & 1.30 & 31.6 & 15.8 & 493 & 1.06 & 1.5
   & 3.9\\
   \enddata
\end{deluxetable*}

\subsection{RFI flagging in the time-channel plane - TCflag}

RFI may be identified in the residual visibilities in the time-channel (TC)
plane of individual baselines.  However, the residual TC plane often has
multi-component sinusoidal patterns caused by RFI or by improper subtraction of
(strong) sources due to several reasons (e.g.  an azimuthally asymmetric antenna
primary beam, time dependent pointing error, uncorrected gain fluctuations
etc.). These visibility fringes tend to inflate estimates of the standard
deviation thereby increasing the threshold above which RFI (localised in both
time and frequency) can be detected.  Therefore we devised a procedure to
eliminate these fringes prior to estimating the sigma threshold for flagging
RFI. The scheme is as follows:

\begin{figure*}
   \gridline{
      \fig{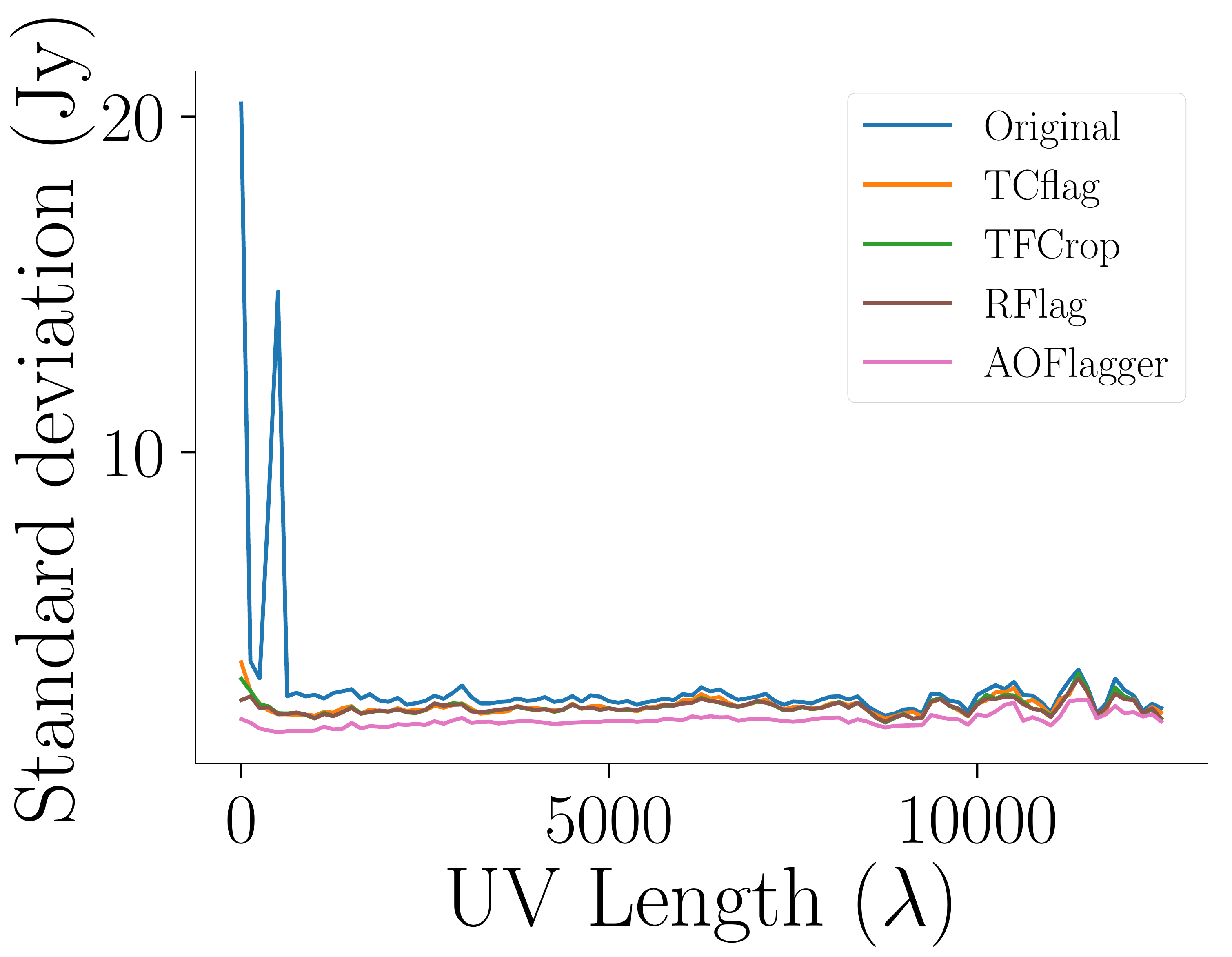}{0.3\linewidth}{3C286}
      \fig{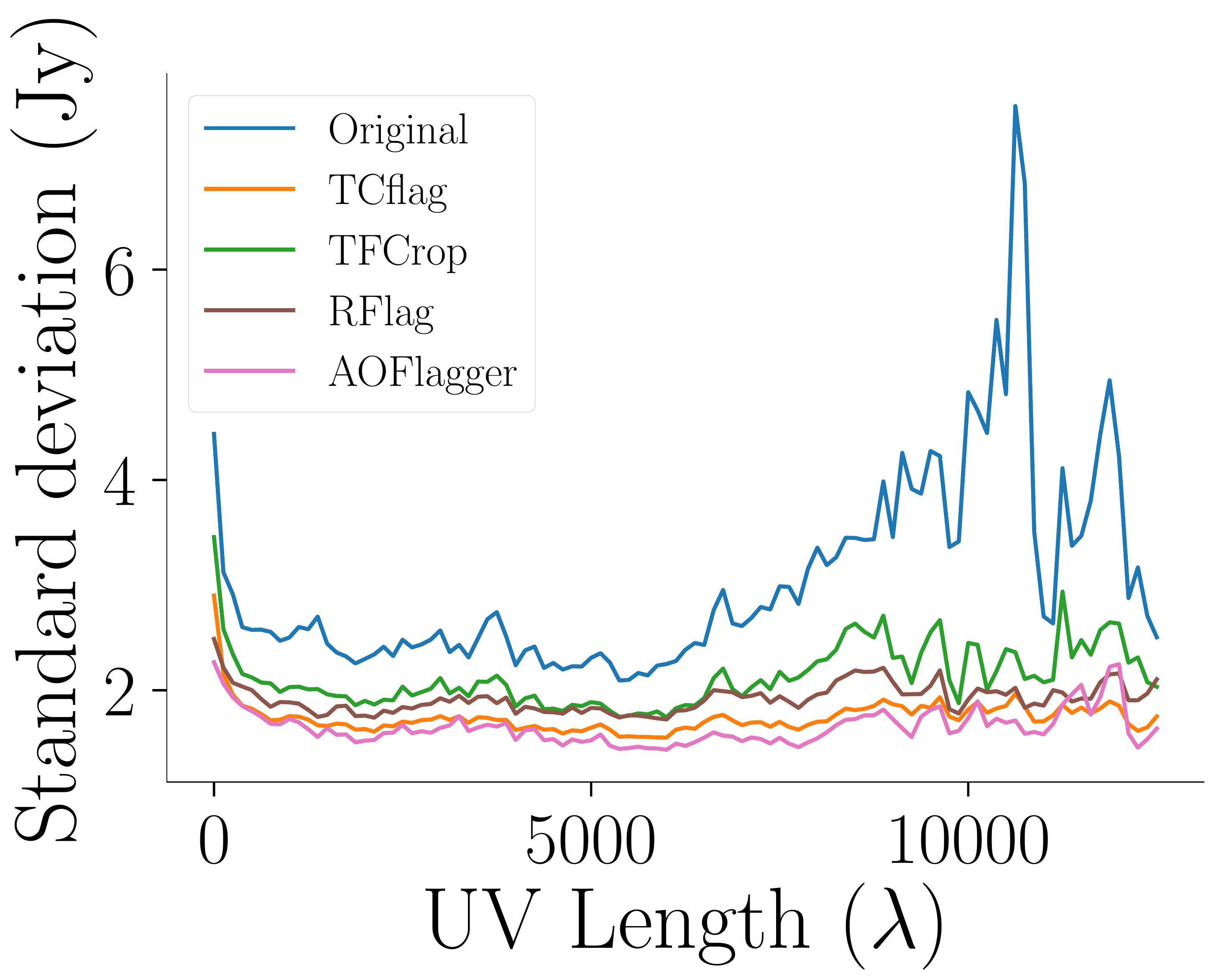}{0.3\linewidth}{VIRMOSC}
      \fig{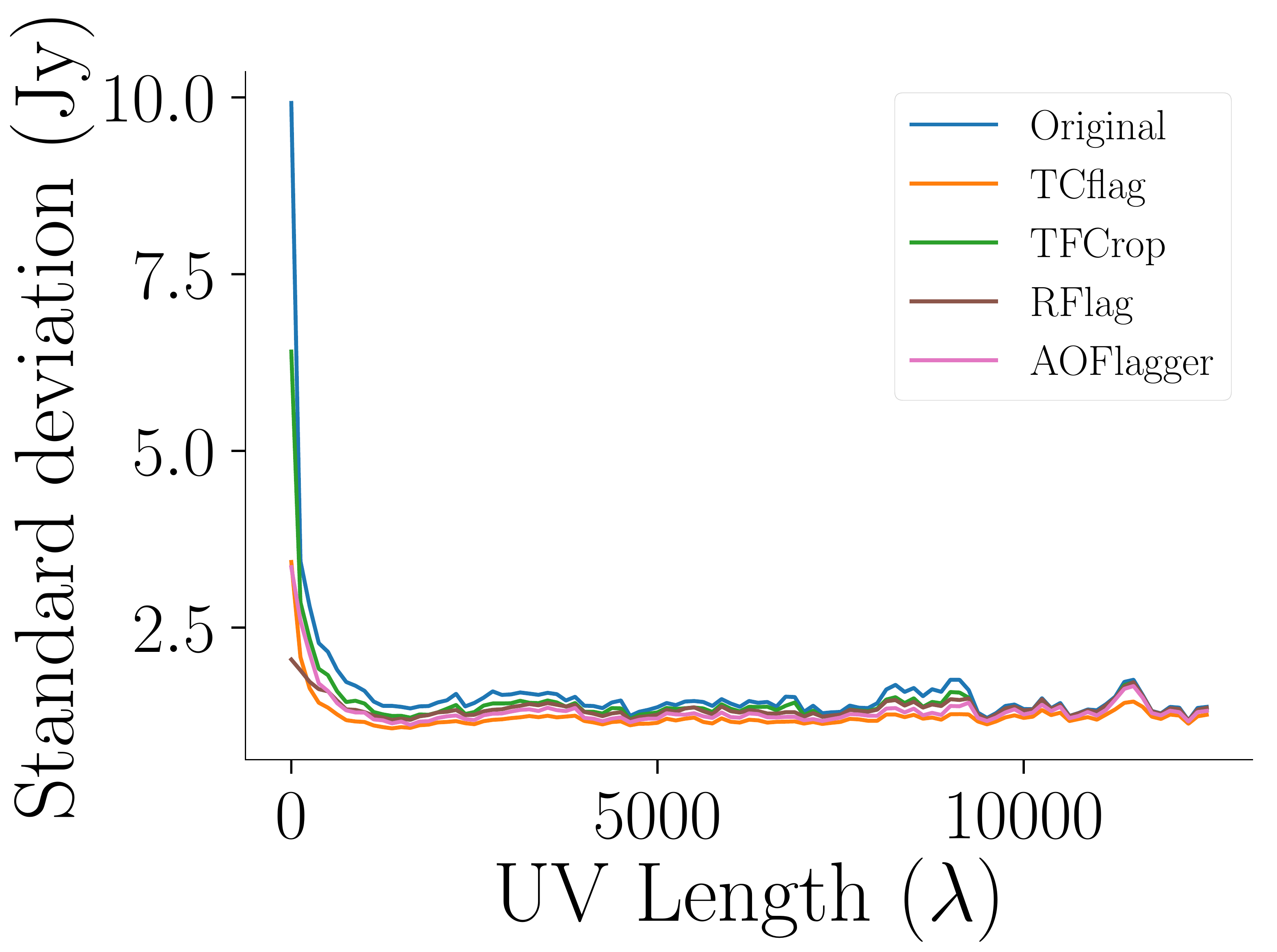}{0.3\linewidth}{J1453+3308}
   }
   \gridline{
      \fig{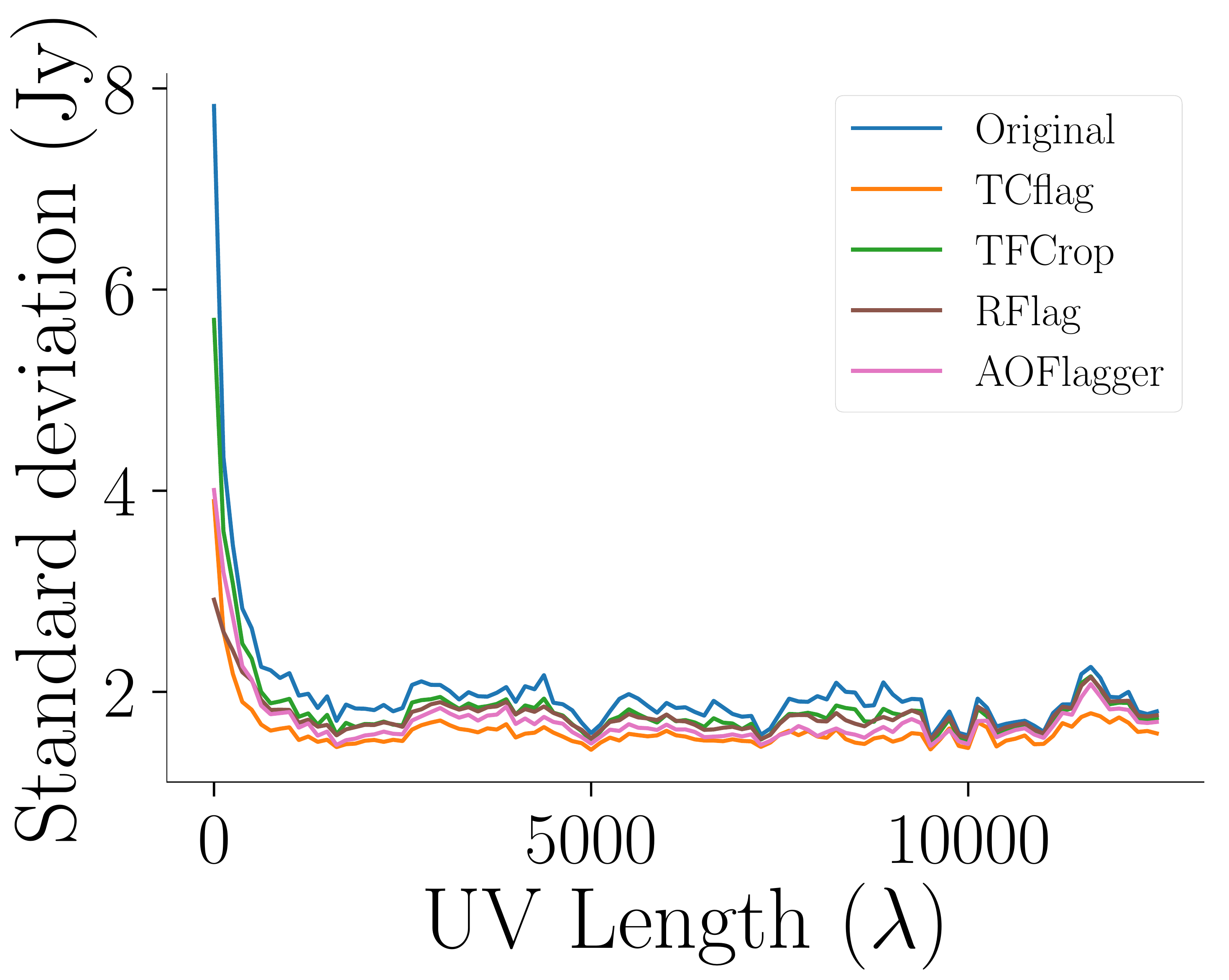}{0.3\linewidth}{J1158+2621}
      \fig{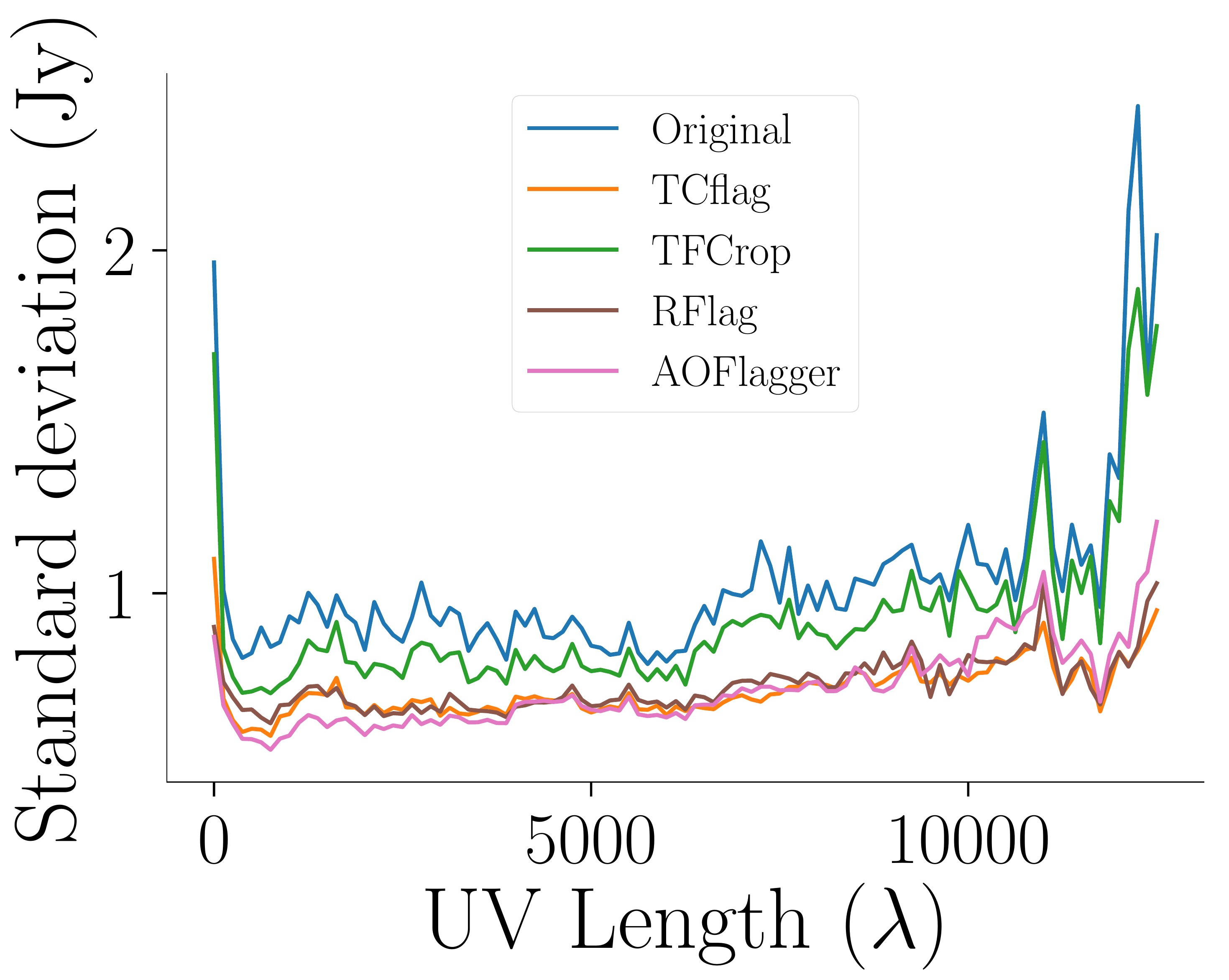}{0.3\linewidth}{A2163}
   }
\caption{A comparison of the time-channel plane RFI flagging effectiveness of
different algorithms. The plots show that TCFlag is competitive with the rest
while using a different procedure to estimate the true background noise in the
time-channel plane of a baseline.}
\label{fig:tc_rms_plots}
\end{figure*}

\begin{enumerate}

   \item For each baseline and polarization take a 2-dimensional Fourier
      transform of the TC plane within a window to obtain the group-delay
      - delay-rate (GD-DR) plane. This window has to be large enough to cover
      a substantial fraction of the fringe period while being smaller than the
      period over which the amplitude of the fringe may vary.

   \item Iteratively sigma clip all components above a threshold in the
      GD-DR plane thereby eliminating the corresponding fringes in the TC
      plane.

   \item Inverse Fourier transform to obtain the fringe-free TC plane.

   \item Identify RFI-affected data using any threshold algorithm (e.g. sigma
      clipping) in the fringe-free TC plane.

   \item Apply the flags to the original data, and restart the process of
      imaging and self-calibration.

\end{enumerate}

This procedure works because the signatures of source structure and localised
RFI differ in the residual TC and GD-DR planes. Any RFI which is localized in
time and frequency will be dispersed over the GD-DR plane, whereas a sinusoidal
fringe due to source structure will show up as compact peaks in GD-DR.

Since the fringes in the residual visibilities arise primarily from incorrectly
subtracted sources, we tried window sizes of 5 - 20 minutes with success. We
settled on a window size of 10 minutes for all sources since it also matched the
scan breaks in our data. This value does not need to be finely tuned.  If the
fringe were to change in amplitude or frequency this will smear the Fourier
signal over several pixels resulting in lower efficiency of RFI detection.
However, one can also run this procedure using several windows sizes in
decreasing succession. This method aims to remove the fringes associated
with improperly subtracted sources in the residual visibilities. The highly
efficient, modern FFT algorithms work well for all window sizes; the process of
FFT and it's inverse results in discrepancies only of the order of double
precision computer numbers ($\sim 10^{-12}$).

Finally, this algorithm can in principle be applied at any stage of image
processing.  Even the presence of real source fringes in time-channel
data will not result in artefacts, as we only modify the flags of the original,
raw data.

\subsection{Observations and Parameters}

All the observations were done with a bandwidth of 16 MHz in the 150 MHz band
and a spectral resolution of 62.5 or 125 kHz. The data was recorded with an
integration time of 2s. The flow of analysis is shown in Figure
\ref{fig:flowchart}.

We used the following parameters in IPFLAG:
\begin{enumerate}
   \item UV-bin size (GRIDflag): 10$\lambda$, from the field of view at 150 MHz.
   \item Smoothing window for median visibility background (GRIDflag):
      5$\times$5 bins.
   \item UV-bin annuli width (GRIDflag): 3, 3 and 6.5 k$\lambda$
   \item Fourier transform window size (TCflag): bandwidth $\times$ 10 min.
   \item Fourier peak detection threshold (TCflag): 3 $\times$ RMS noise
   \item RFI threshold (both): 3 $\times$ RMS noise
\end{enumerate}

We transferred the RFI flags from IPFLAG to the raw data, and repeated
the entire process of imaging, self-calibration, residual flagging, and flag
transfer. Finally, the data was again self-calibrated and imaged for the final
result.  (see Figure \ref{fig:flowchart}). The final image covers $6.25\degr
\times 6.25 \degr$ with a pixel size of $4.5\arcsec$. The sources were extracted
from the images  using the PyBDSF source finder \citep{mohan_pybdsf_2015},
running with identical parameters across all images.

Stand-alone calibration using the standard CASA/AIPS recipes resulted in
flux density scale errors of up to 15\%. This becomes important while comparing
the absolute noise in an image though it does not affect the relative change in
noise with the application of RFI flagging procedures. We therefore anchored our
flux density scale to TGSS-ADR \citep{intema2017gmrt} using the sources in
common to the two observations. The fractional discrepancy in source flux
densities is shown in Figure \ref{fig:flux_scale}. For reasons we do not
understand, but which may have to do with something similar to CLEAN bias
\citep{condon1998nrao, cohen2007vla} we find a flux dependent fractional
discrepancy between TGSS and our flux densities. The fractional discrepancy
changes between  -0.5\% to -9\% between sources above and below 100 mJy.

\begin{figure*}
   \gridline{
      \fig{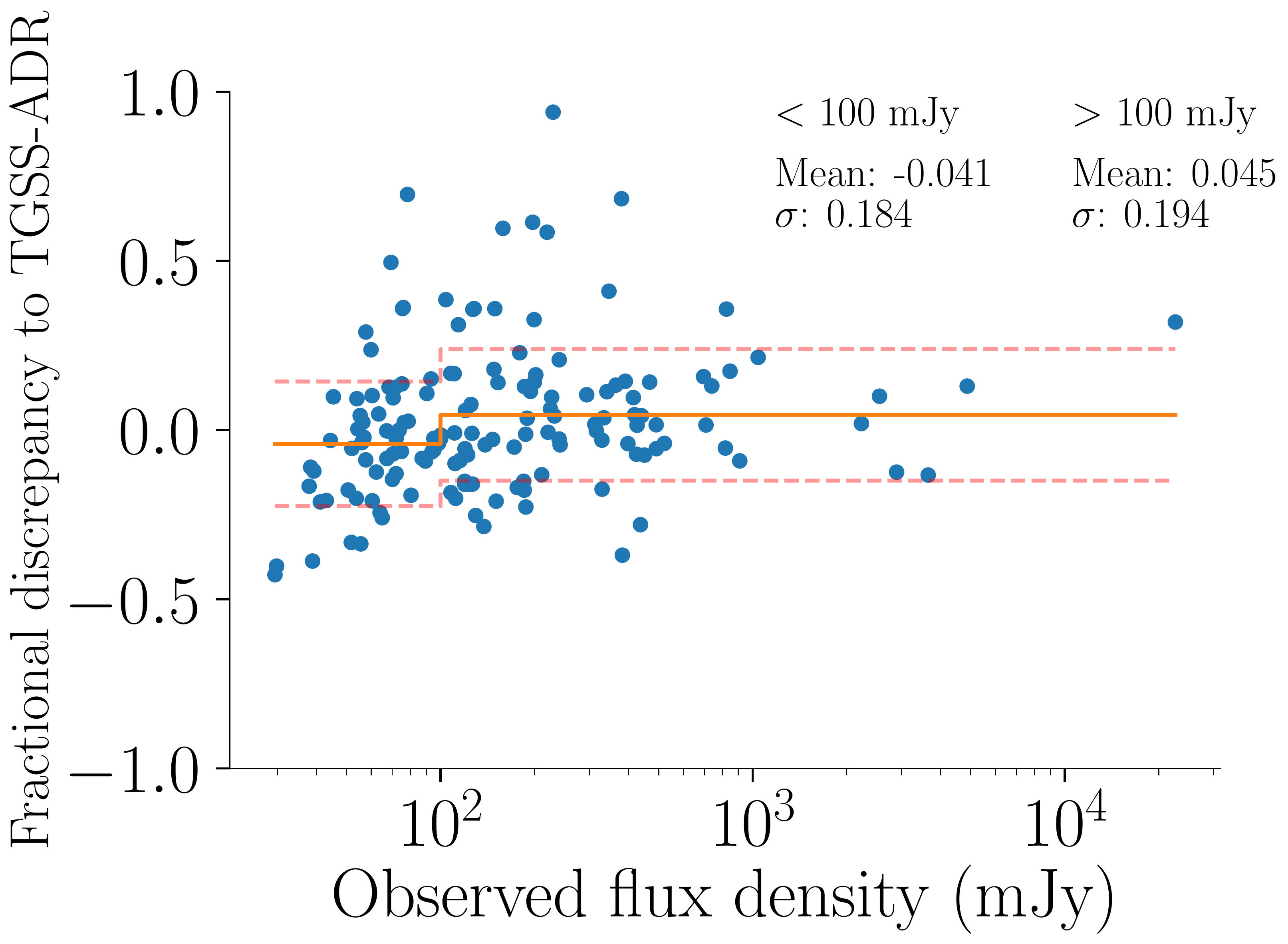}{0.35\linewidth}{3C286}
      \fig{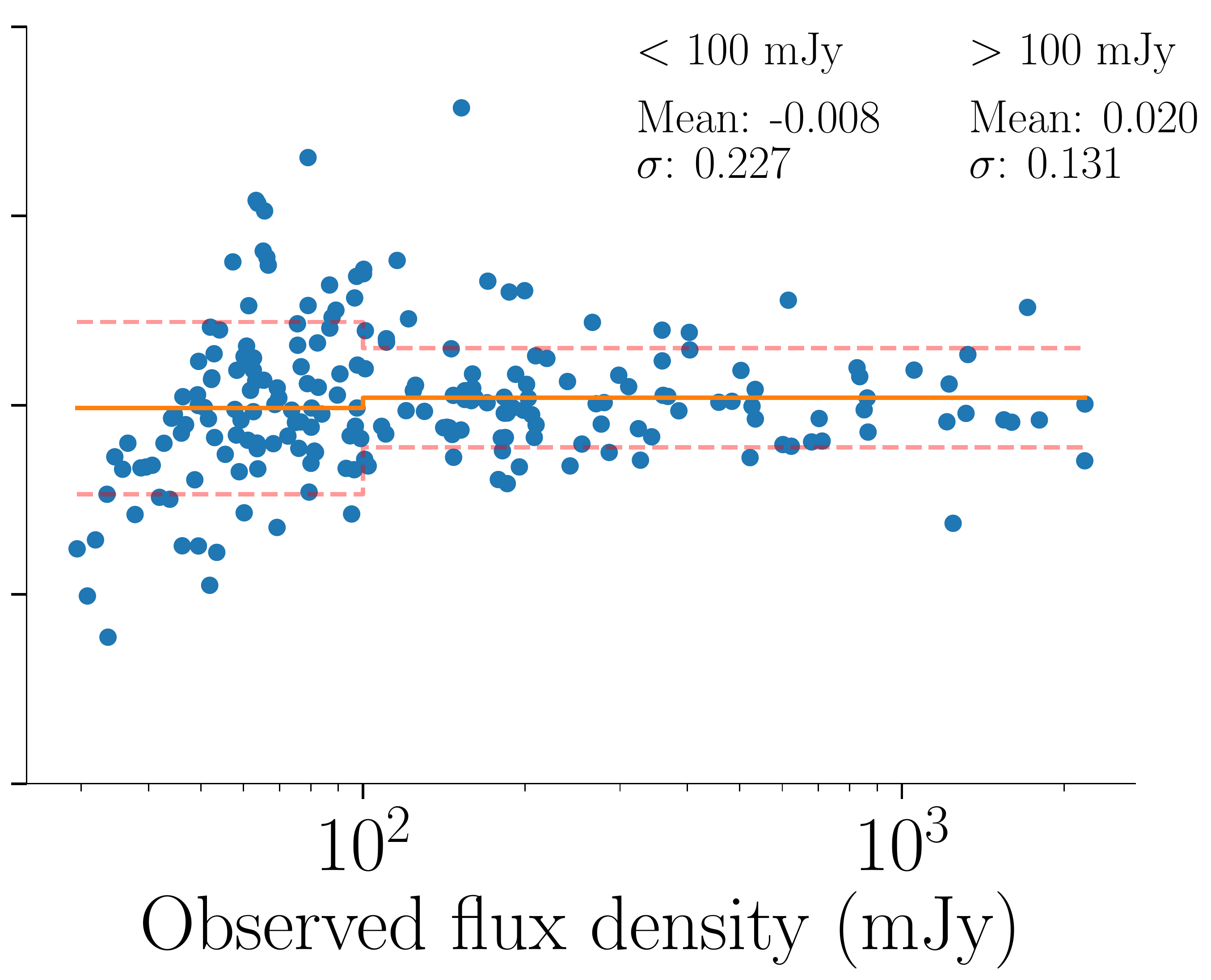}{0.3\linewidth}{VIRMOSC}
      \fig{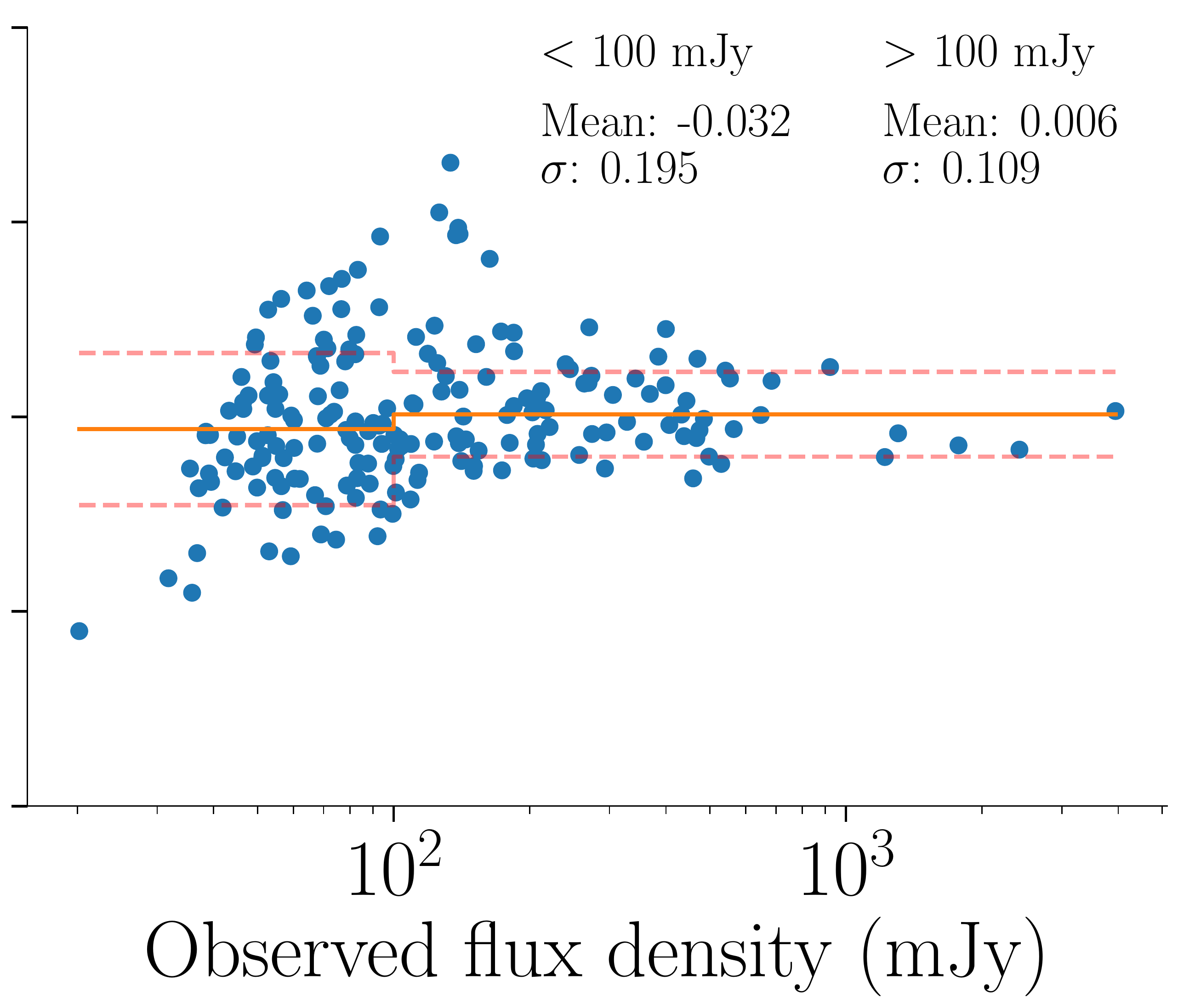}{0.3\linewidth}{J1453+3308}
   }
   \gridline{
      \fig{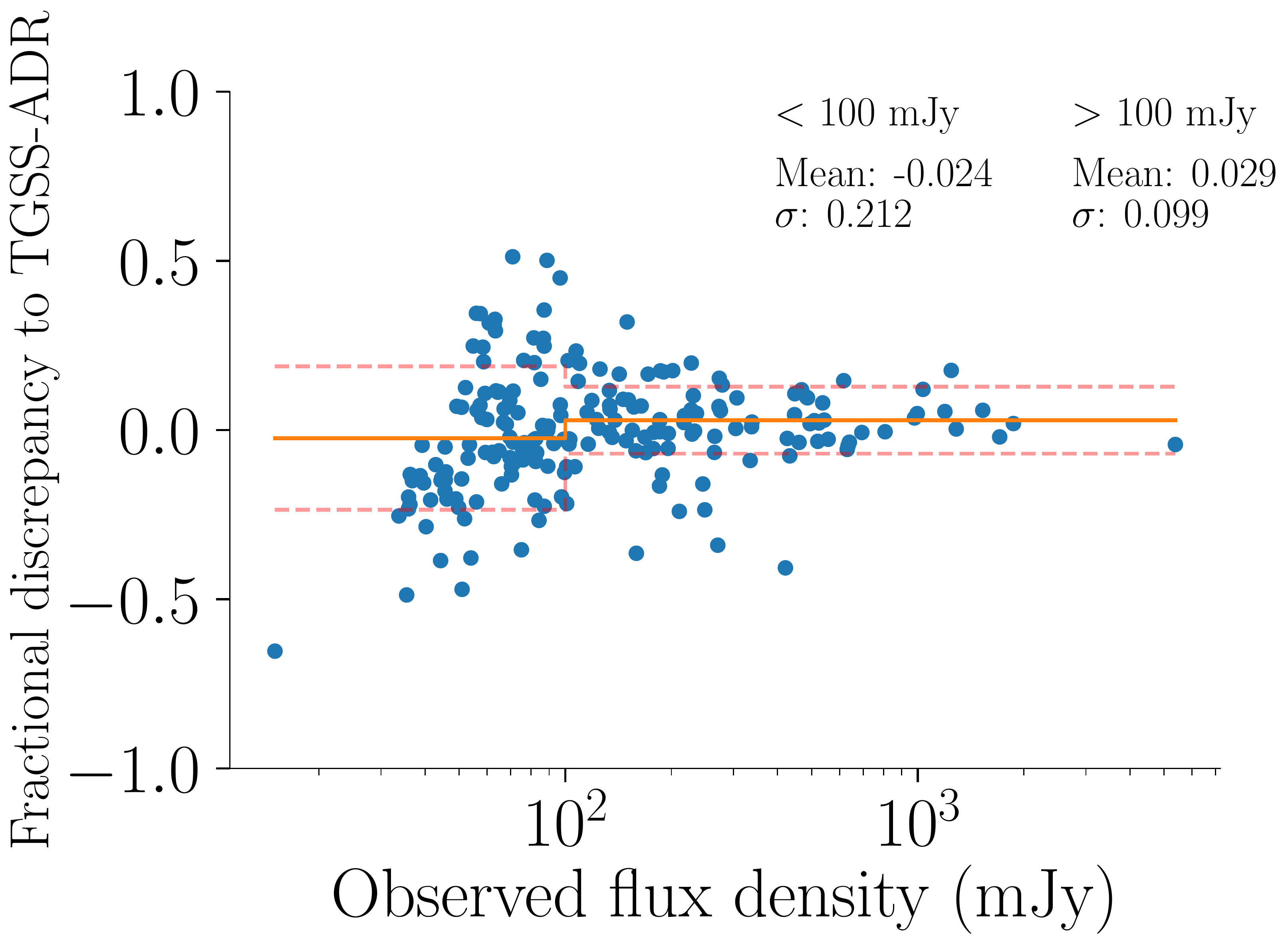}{0.35\linewidth}{J1158+2621}
      \hfill
      \fig{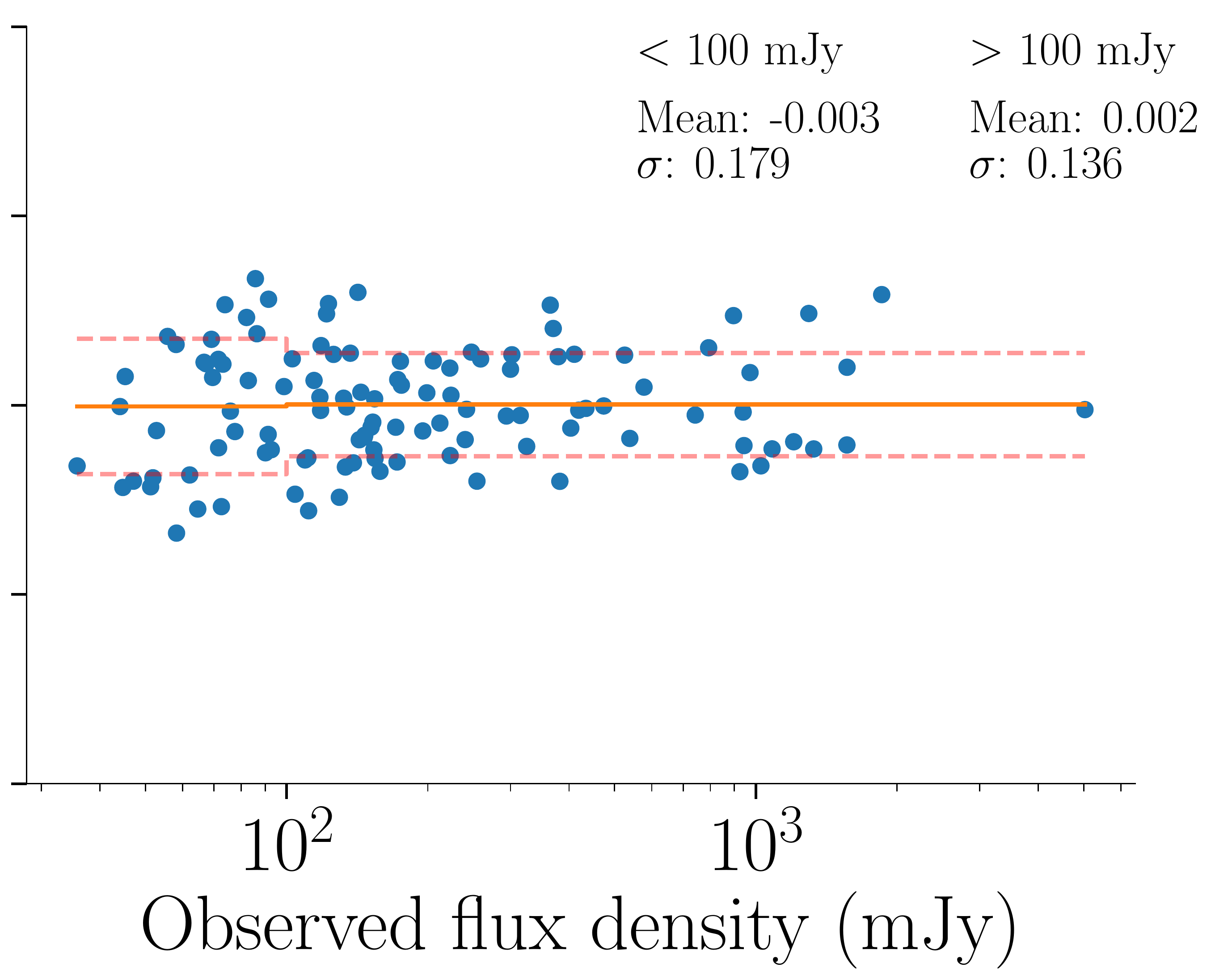}{0.3\linewidth}{A2163}
   }
   \caption{The plots show the fractional flux density discrepancy between our
   data and TGSS-ADR for point sources common to both. The solid line shows the
   mean fractional discrepancy (corrected to zero by suitable scaling) and the
   dashed lines indicate the standard deviation of the scatter for sources above
   and below 100 mJy.}
   \label{fig:flux_scale}
\end{figure*}

\section{Efficacy}

We applied IPFLAG to real data from the GMRT at 150 MHz and compared the results
with and without the same. The 150 MHz band of the GMRT is important for
a variety of astrophysical phenomena but is under-utilized because of the
presence of strong RFI.  We felt that this comparison using real data would be
a more realistic appraisal of the algorithms than simulations with well-behaved
noise.

We targeted 5 fields - VIRMOSC (GMRT observation code: 14RAA01),
J1453+3308 (27\textunderscore063), J1158+2621 (27\textunderscore063), A2163
(16\textunderscore259), and 3C286 (TGSS data, \citealt{intema2017gmrt}).

\paragraph{3C286} A commonly used flux density calibrator source, which is
compact and has a flux of 26 Jy at 150 MHz. The field is dominated by point
sources with almost no extended emission. However, for reasons that are not
understood this field showed a reduced flux density in TGSS-ADR by $\sim$25\%
\citep[see][]{intema2017gmrt}.

\paragraph{VIRMOSC} A field dominated by point sources. The strongest point
source in the field is $\sim$ 1.7 Jy while it has a single diffuse (4 arcmin)
source of 300 mJy.

\paragraph{J1453+3308 and J1158+2621} Double-double radio galaxies, with
diffuse outer lobes spanning $4-7 \arcmin$.

\paragraph{A2163} This is a galaxy cluster with a $14\arcmin$ low surface
brightness radio halo.\\

Figure \ref{fig:gridded_uv_plane} shows examples of the elimination of the RFI
hotspots from the median binned UV plane. Only the upper half of the
UV plane is shown in the plot; the lower half is simply the Hermitian
conjugate. The band of higher intensity seen at U $\equiv$ [-100 $\lambda$, 100
$\lambda$] (e.g.  J1453+3308 in Figure \ref{fig:gridded_uv_plane}) arises from
the inability of the RfiX algorithm to mitigate RFI in regions where the
fringe-stop frequency is close to zero.

Figure \ref{fig:tc_rms_plots} shows a comparison between TCflag
(described here) and existing flagging tools RFlag, TFCrop (both implemented
in CASA) and AOFlagger \citep{offringa_morphological_2012,
offringa_post-correlation_2010}. Our algorithm TCflag is competitive with the
rest while using a different procedure to estimate the true noise background.

The total data flagged by IPFLAG was 2.4 - 15.7\%, and the corresponding loss in
UV-bins was 1.2 - 3.9\%.  Table \ref{table:image_parameters} shows that the
change in the dirty beam parameters is small before and after IPFLAG confirming
that our procedure did not change the UV coverage despite the loss of
a substantial amount of data. We used the same restoring beam before and after
IPFLAG in all our targets. The post imaging comparisons are plotted in Figure
\ref{fig:all_plots}.

\begin{figure*}[htbp]
   \gridline{ \text{3C286} }
   \gridline{
      \fig{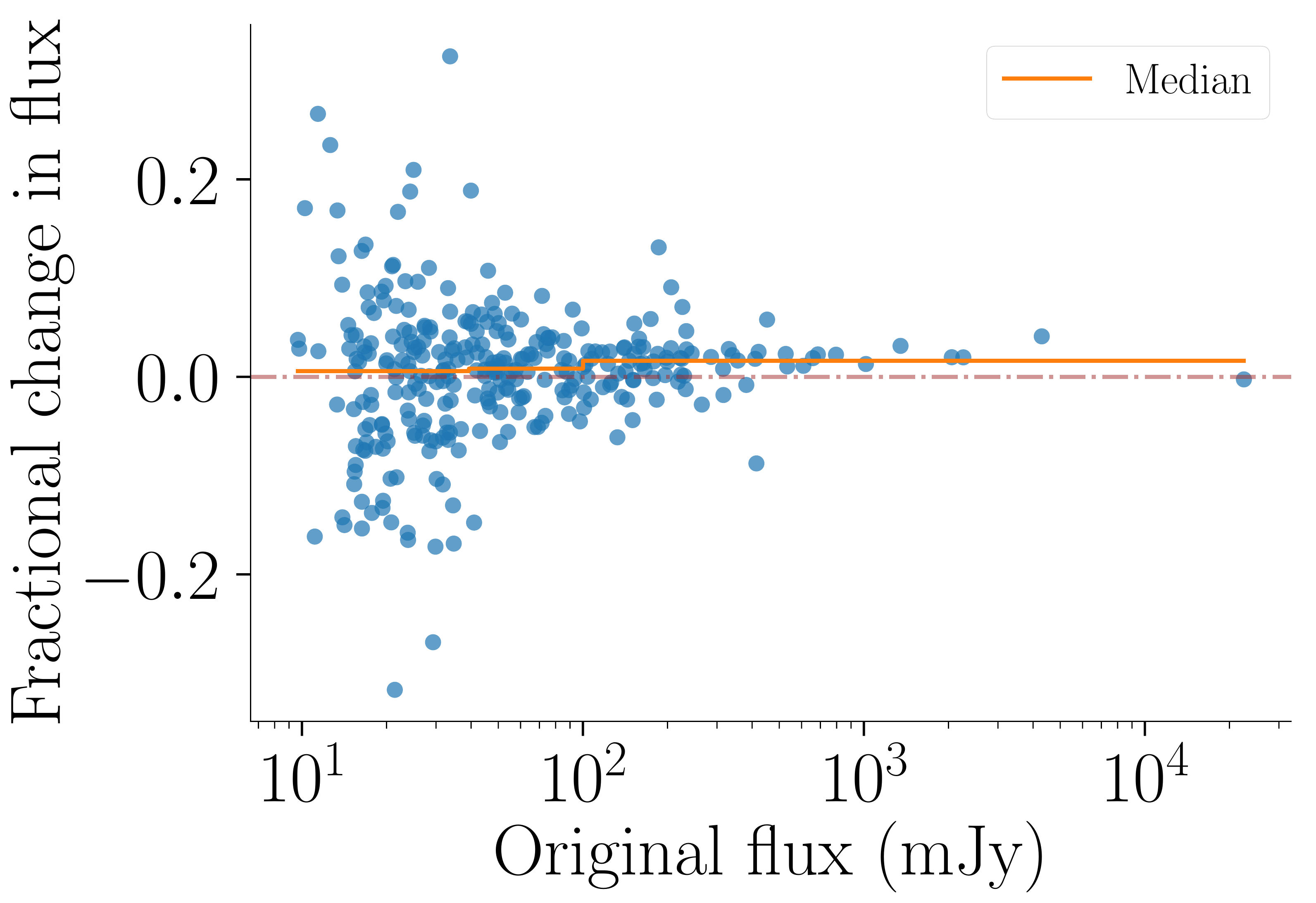}{0.20\linewidth}{(i-a)}
   \fig{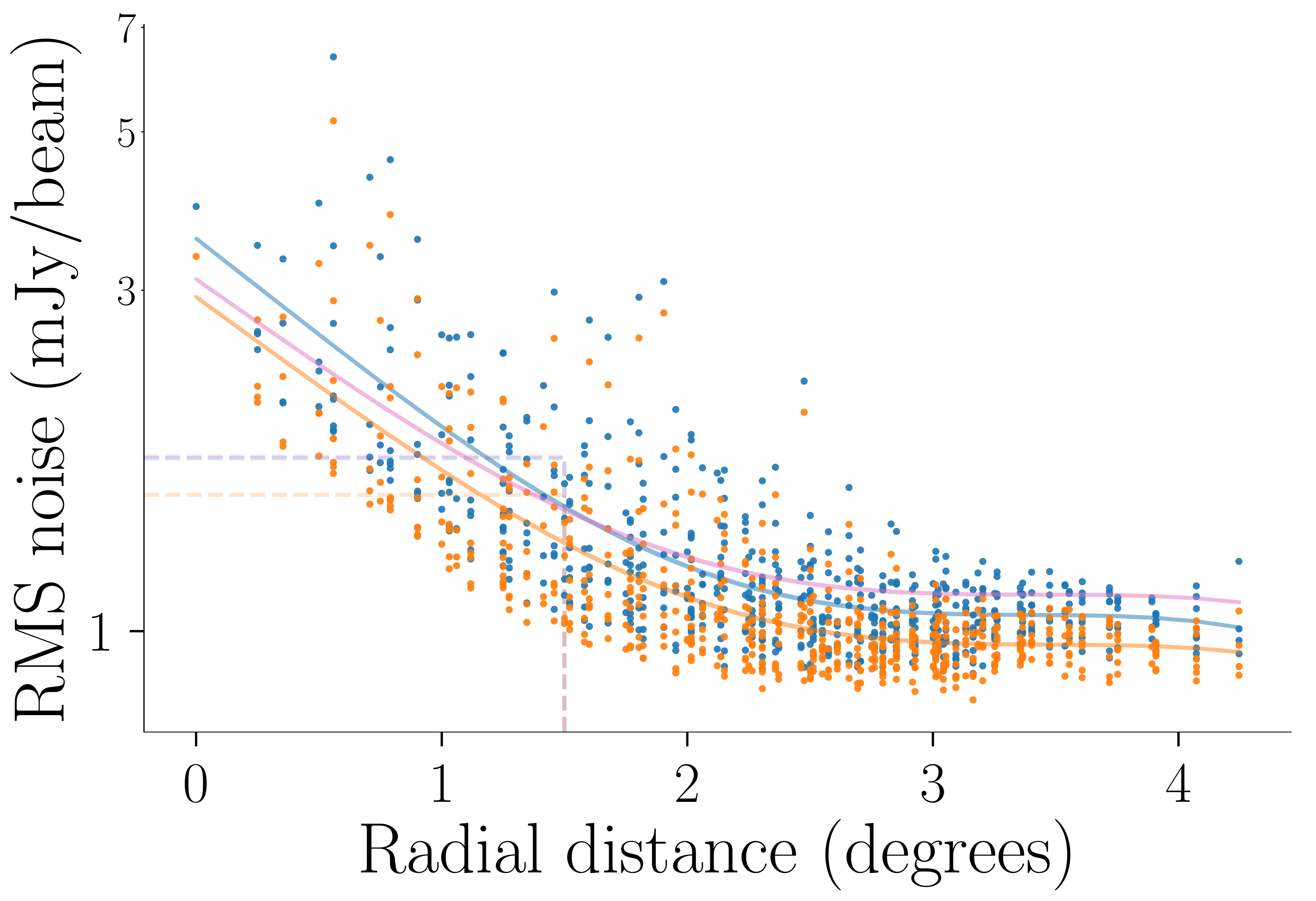}{0.20\linewidth}{(i-b)}
   \fig{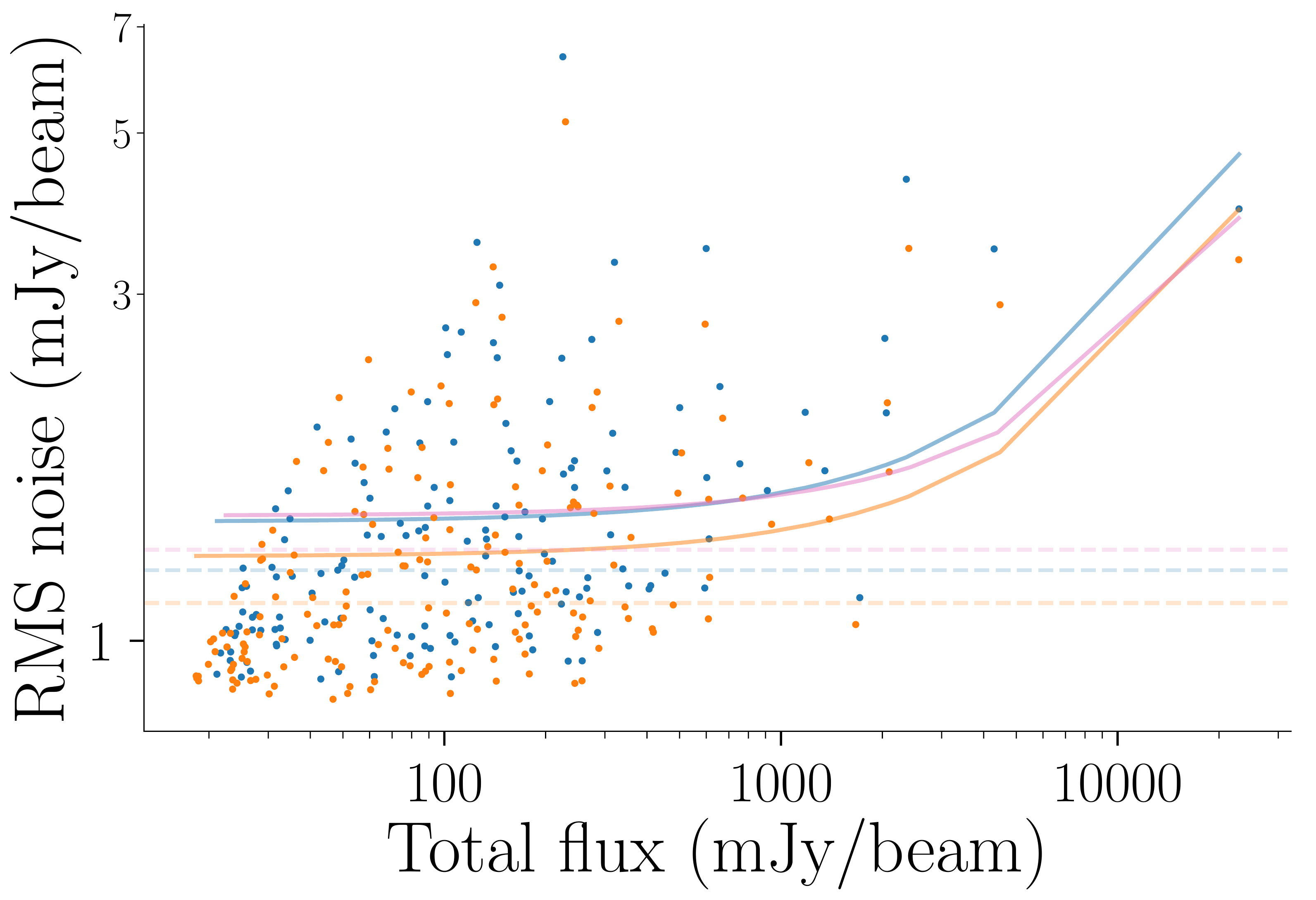}{0.20\linewidth}{(i-c)}
   \fig{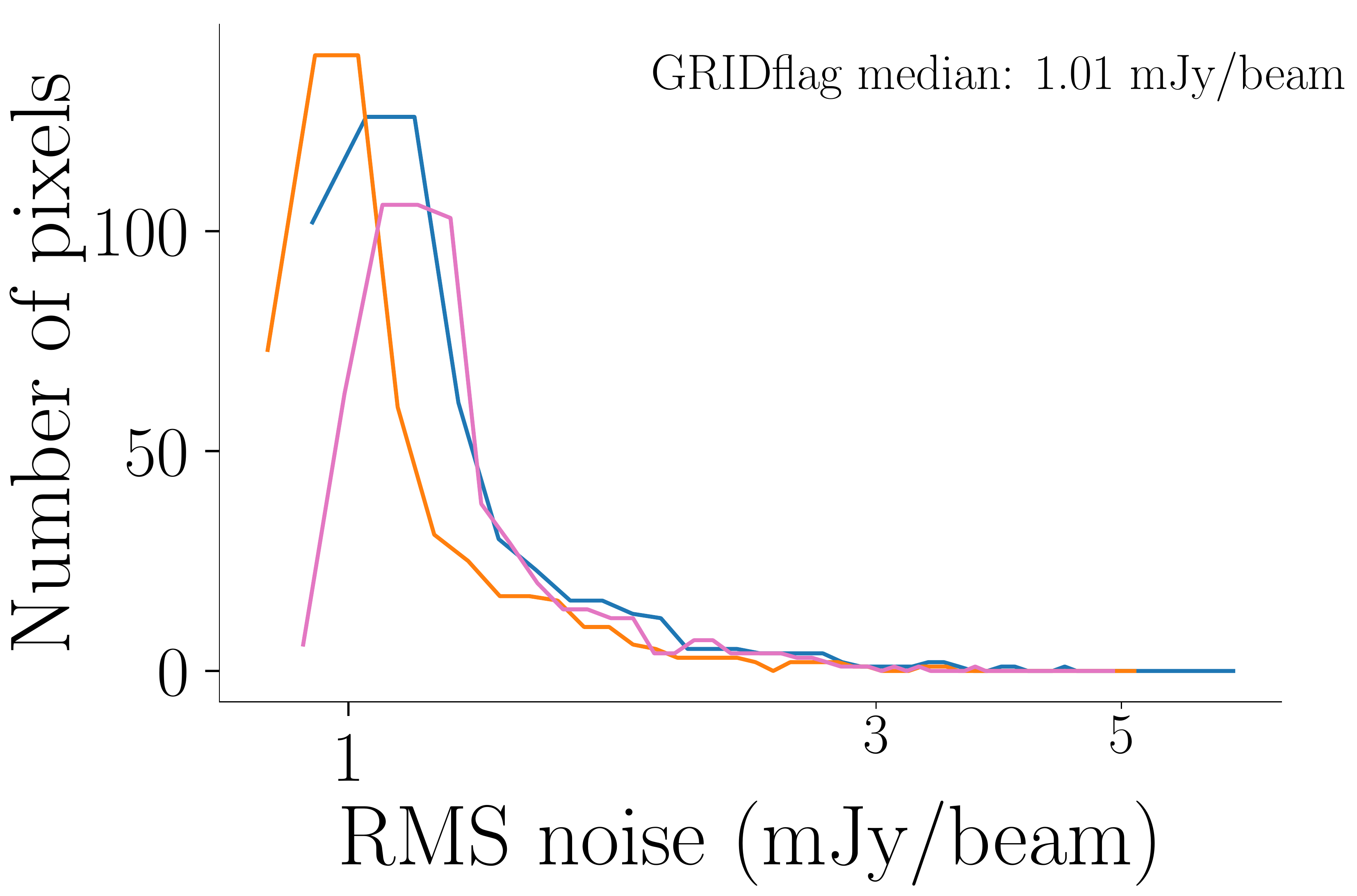}{0.20\linewidth}{(i-d)}
   \fig{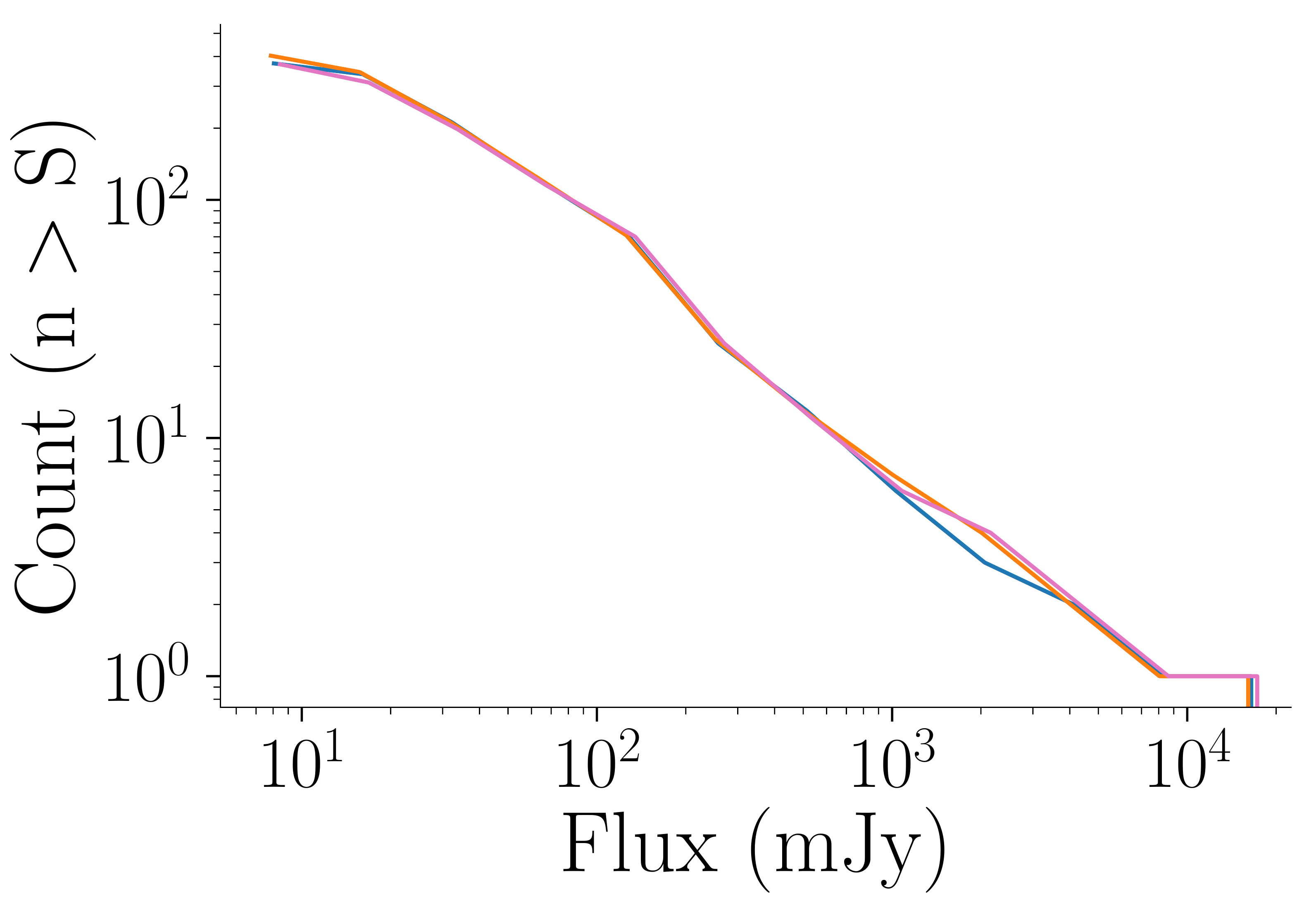}{0.20\linewidth}{(i-e)}
}
   \gridline{
      \text{VIRMOSC}
   }
   \gridline{
      \fig{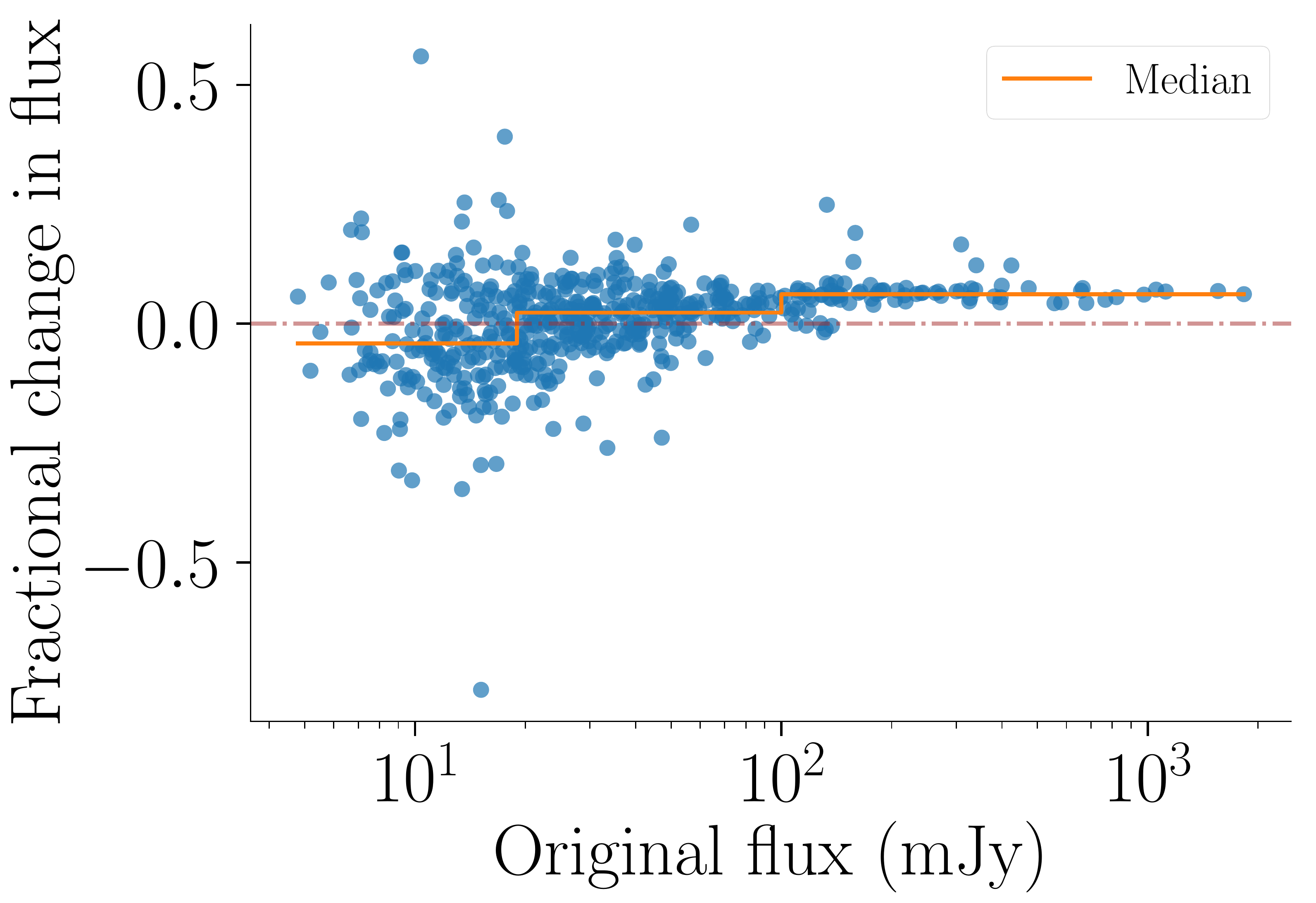}{0.20\linewidth}{(ii-a)}
   \fig{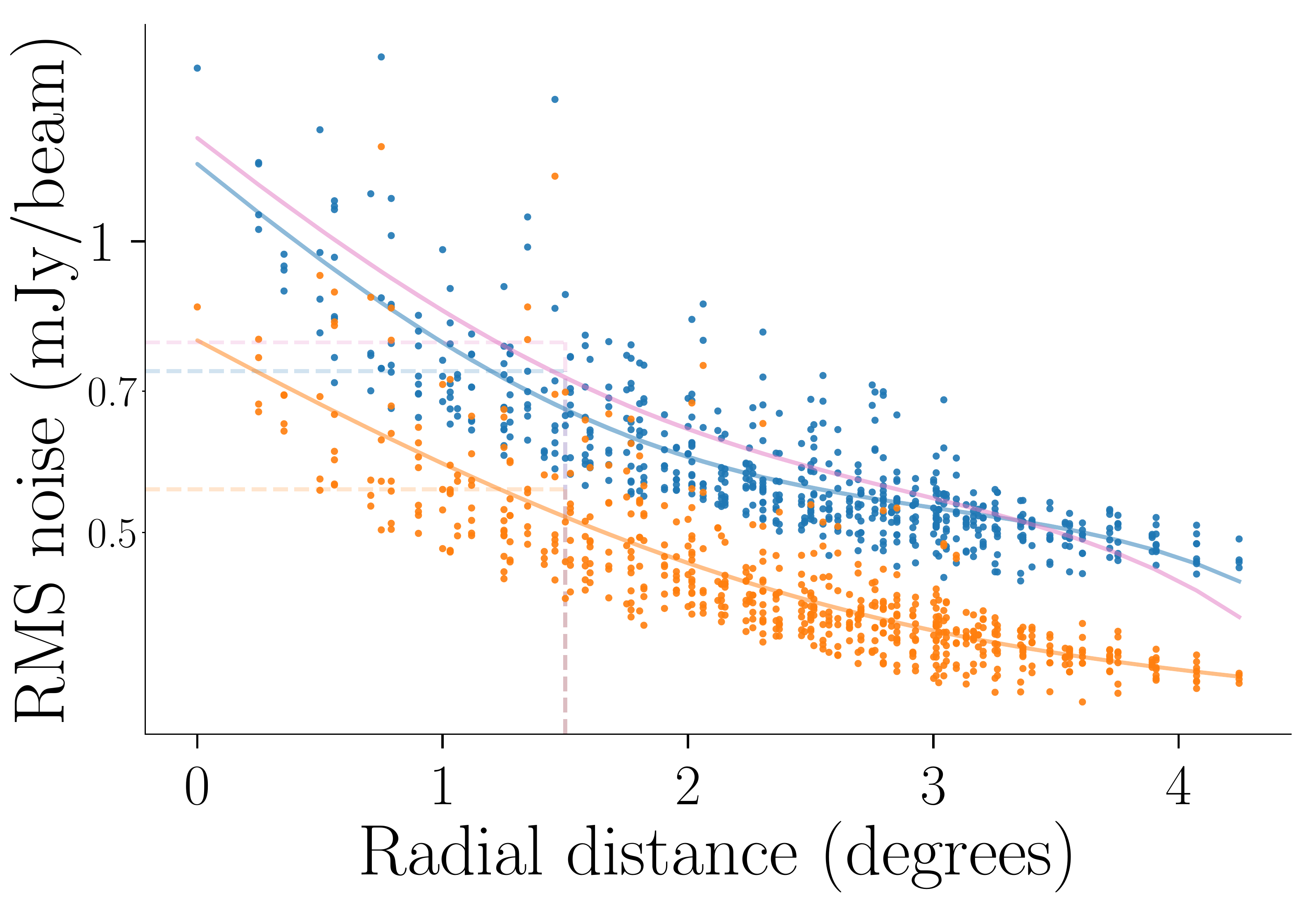}{0.20\linewidth}{(ii-b)}
   \fig{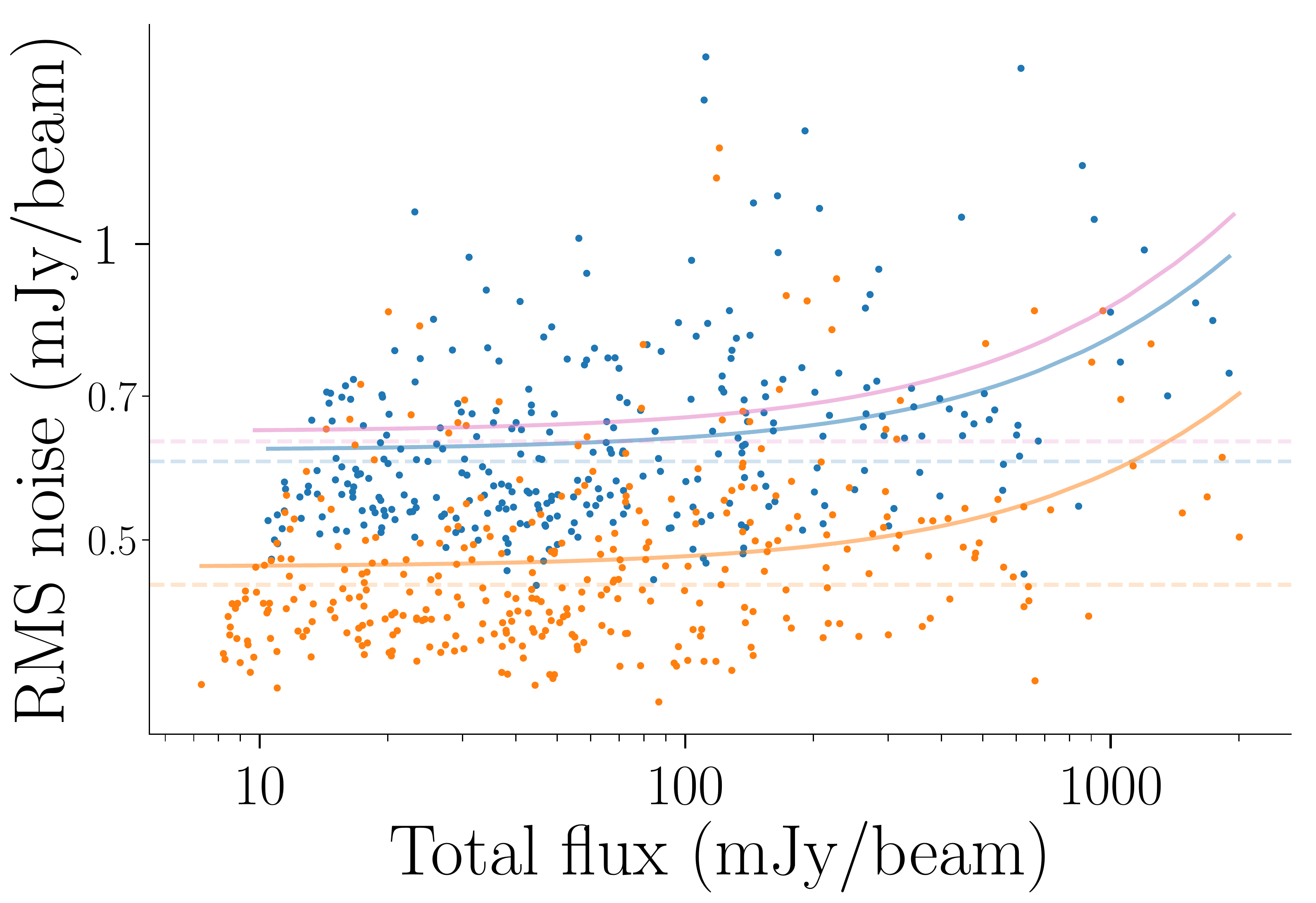}{0.20\linewidth}{(ii-c)}
   \fig{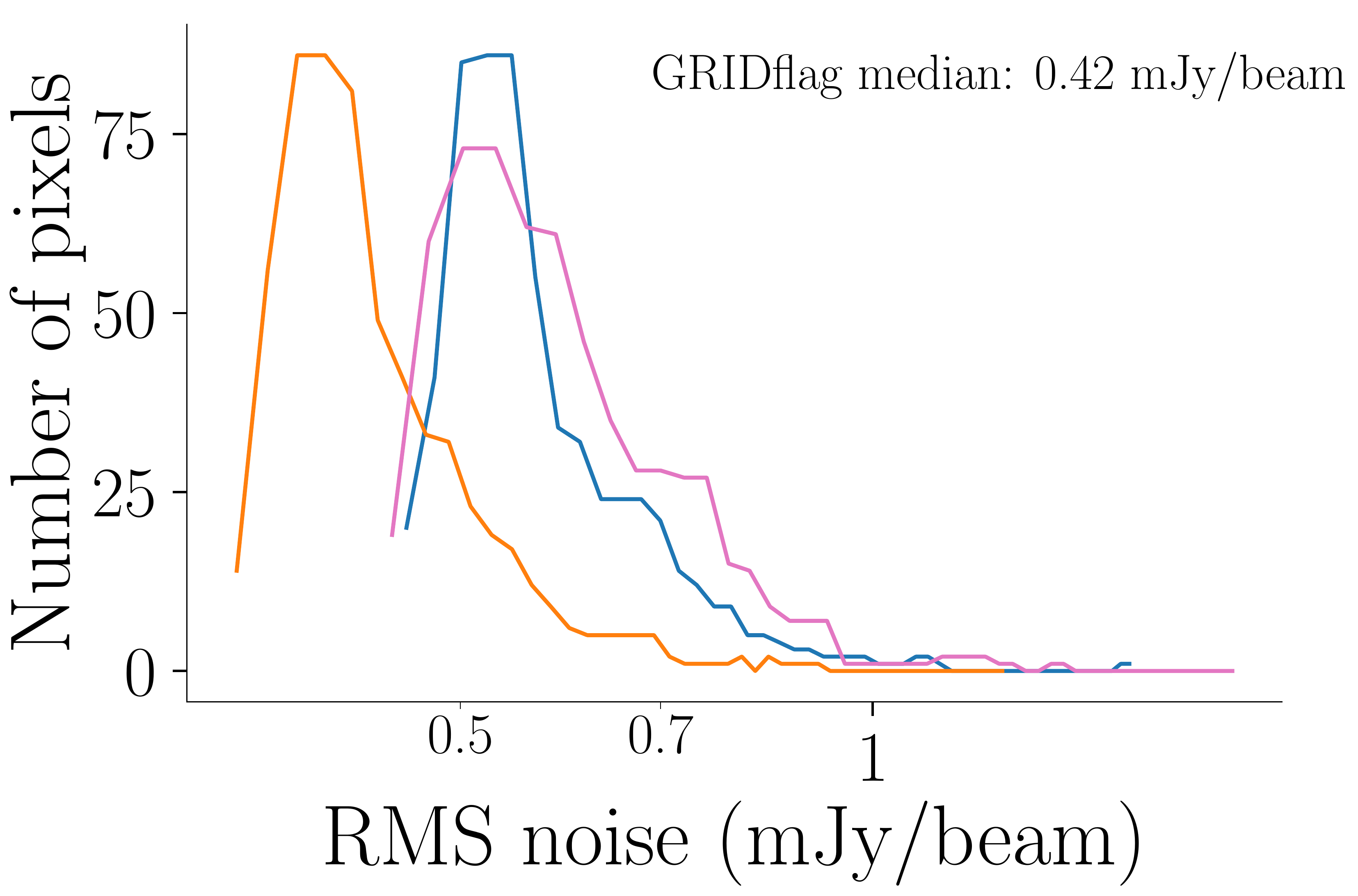}{0.20\linewidth}{(ii-d)}
   \fig{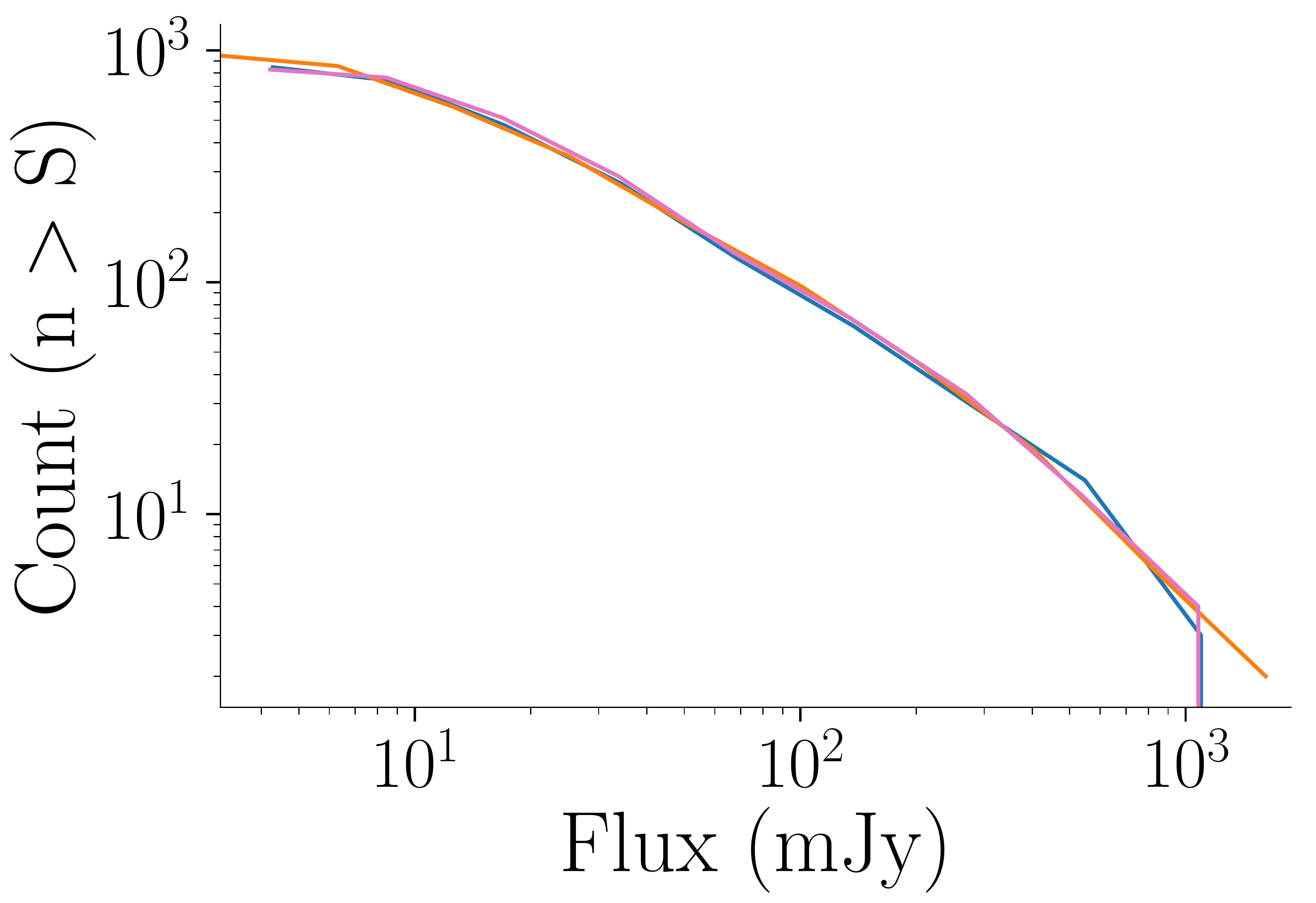}{0.20\linewidth}{(ii-e)}
}
   \gridline{
      \text{J1453+3308}
   }
   \gridline{
      \fig{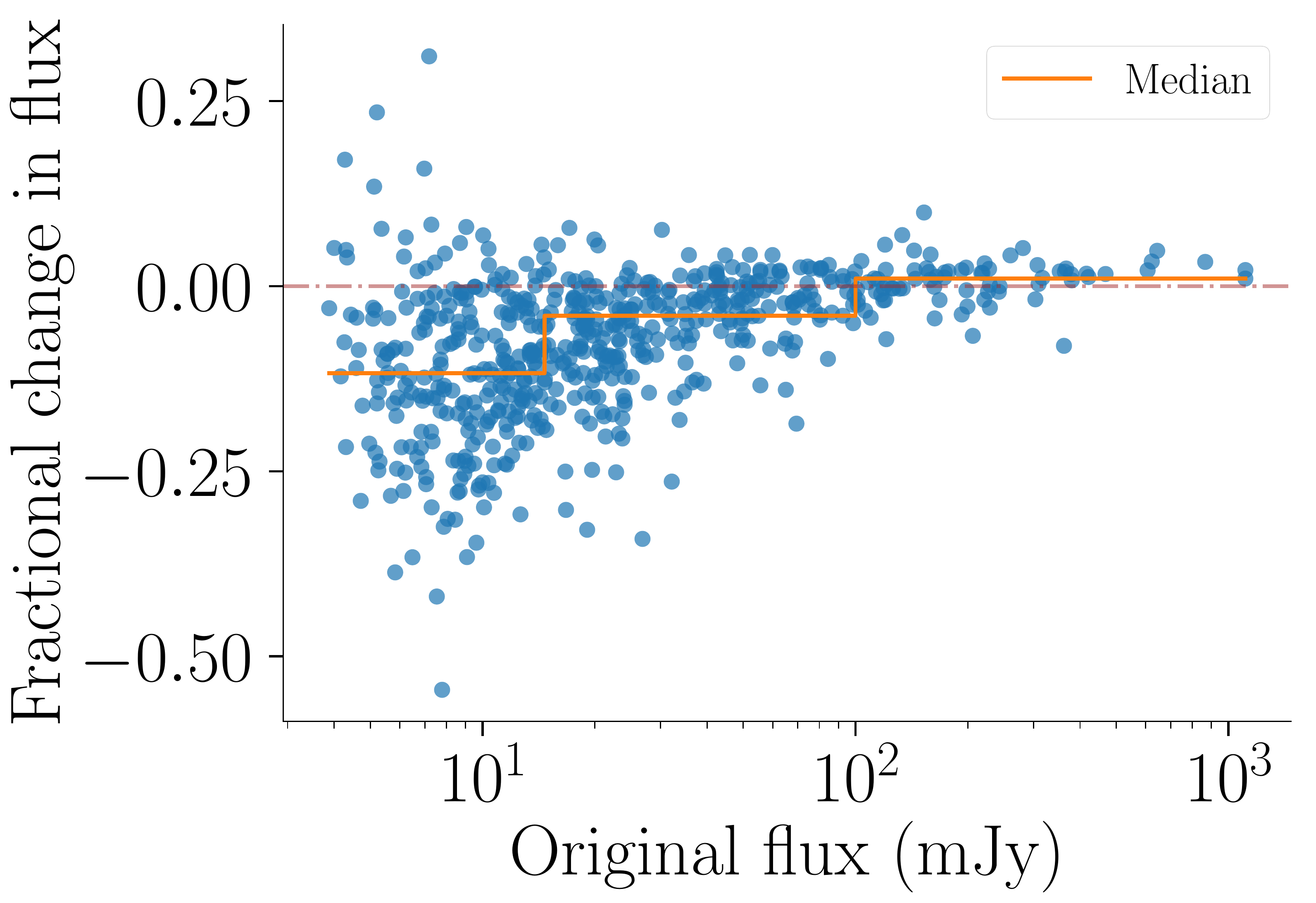}{0.20\linewidth}{(iii-a)}
   \fig{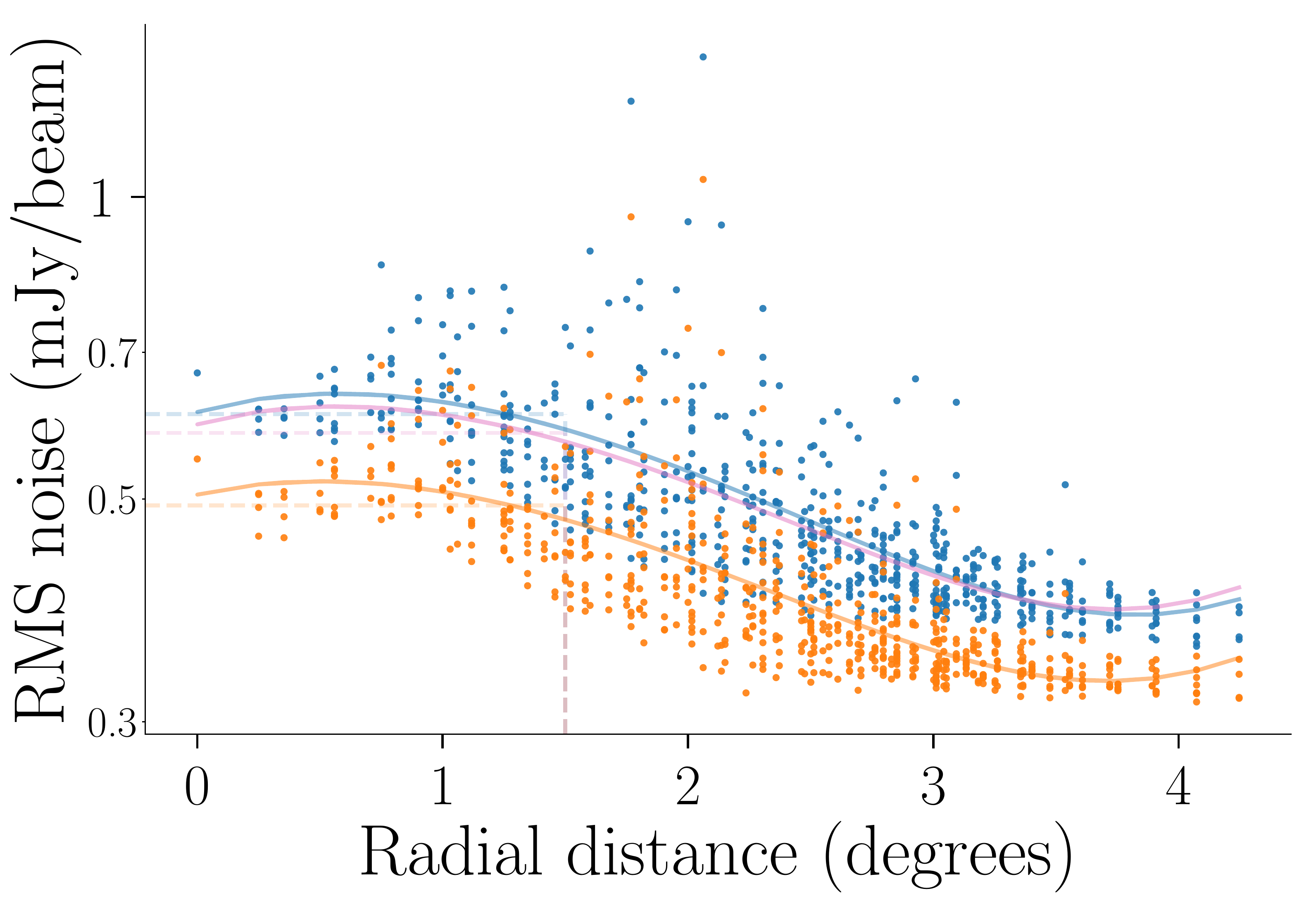}{0.20\linewidth}{(iii-b)}
   \fig{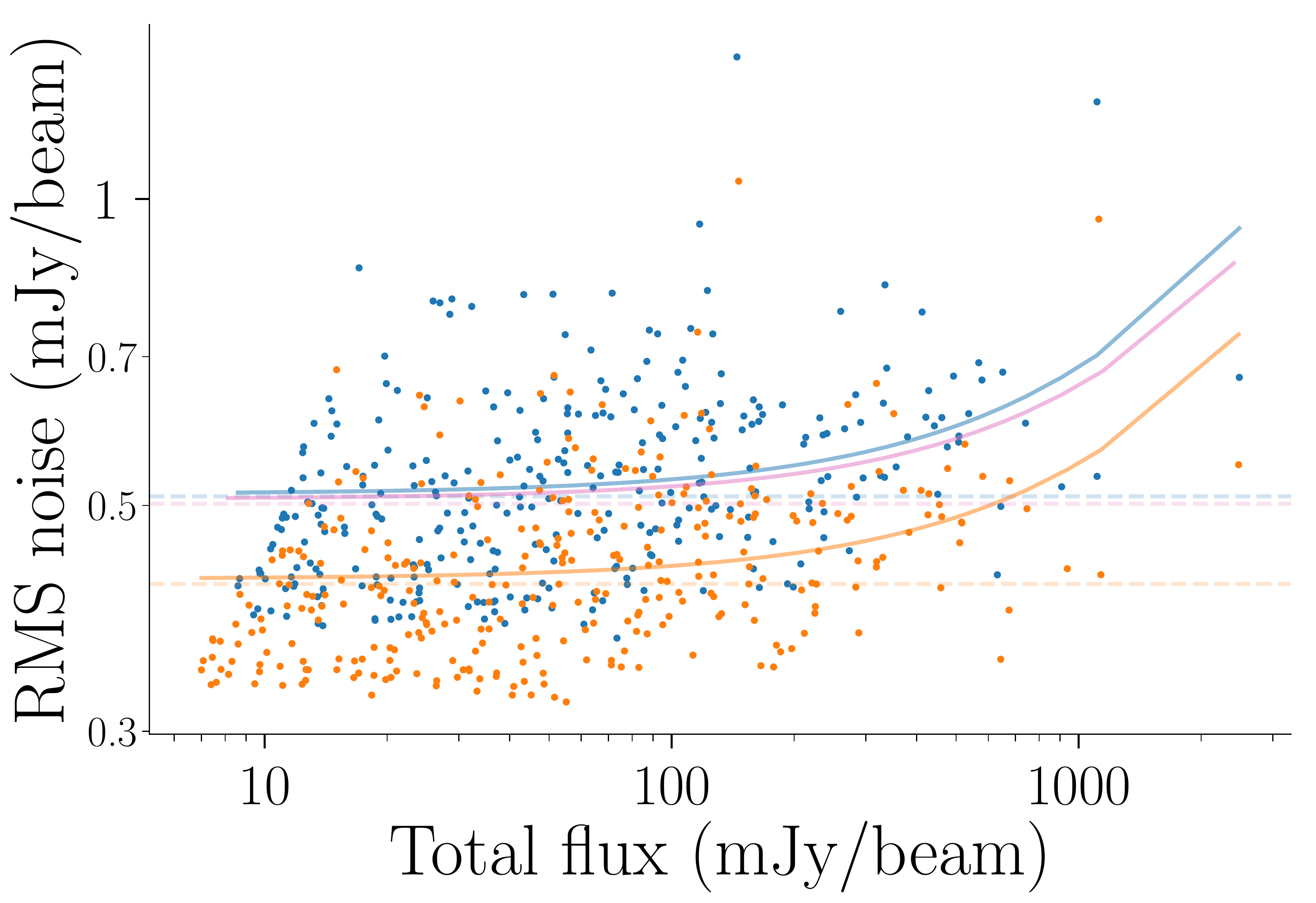}{0.20\linewidth}{(iii-c)}
   \fig{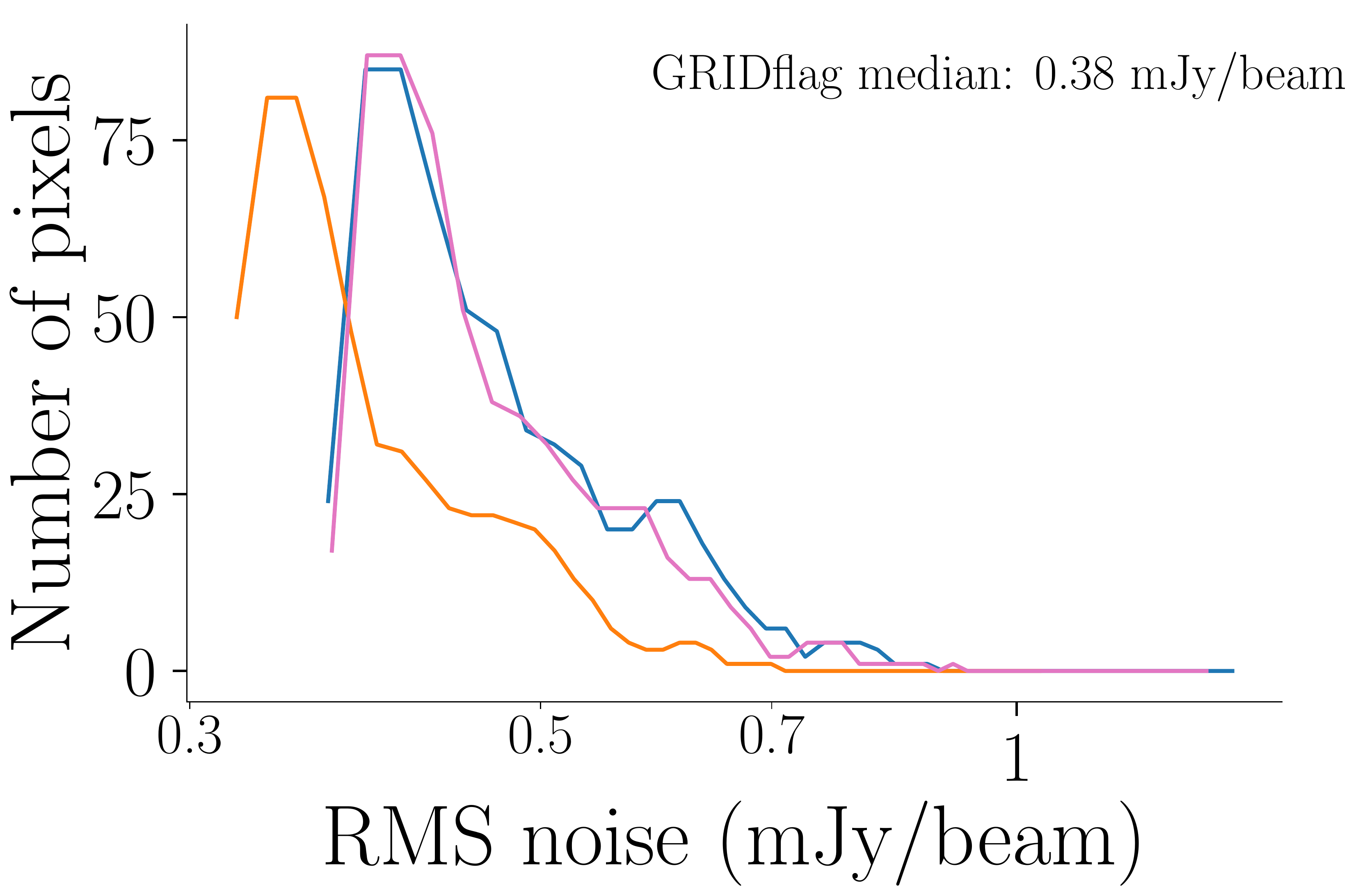}{0.20\linewidth}{(iii-d)}
   \fig{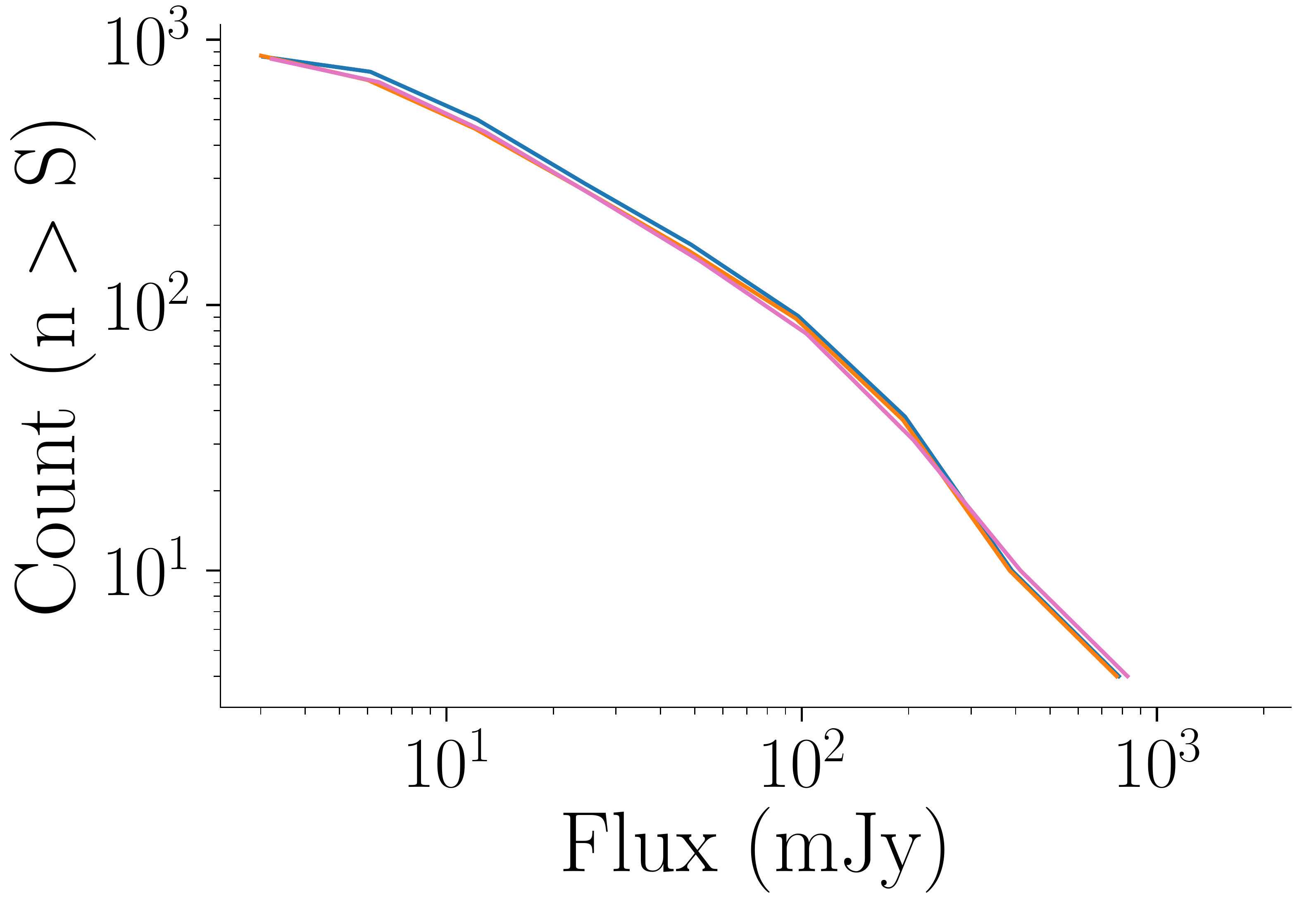}{0.20\linewidth}{(iii-e)}
}
   \gridline{
      \text{J1158+2621}
   }
   \gridline{
      \fig{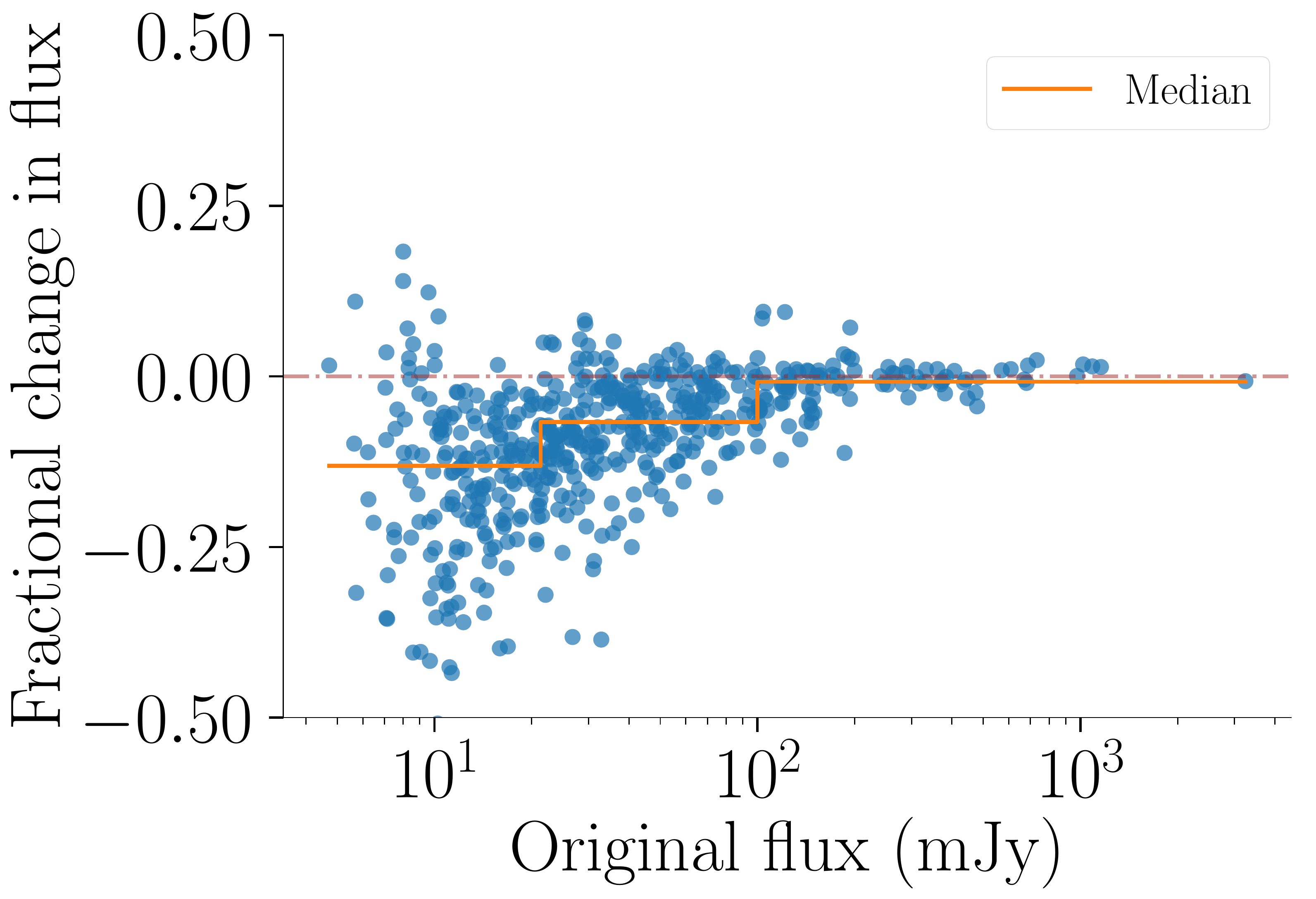}{0.20\linewidth}{(iv-a)}
   \fig{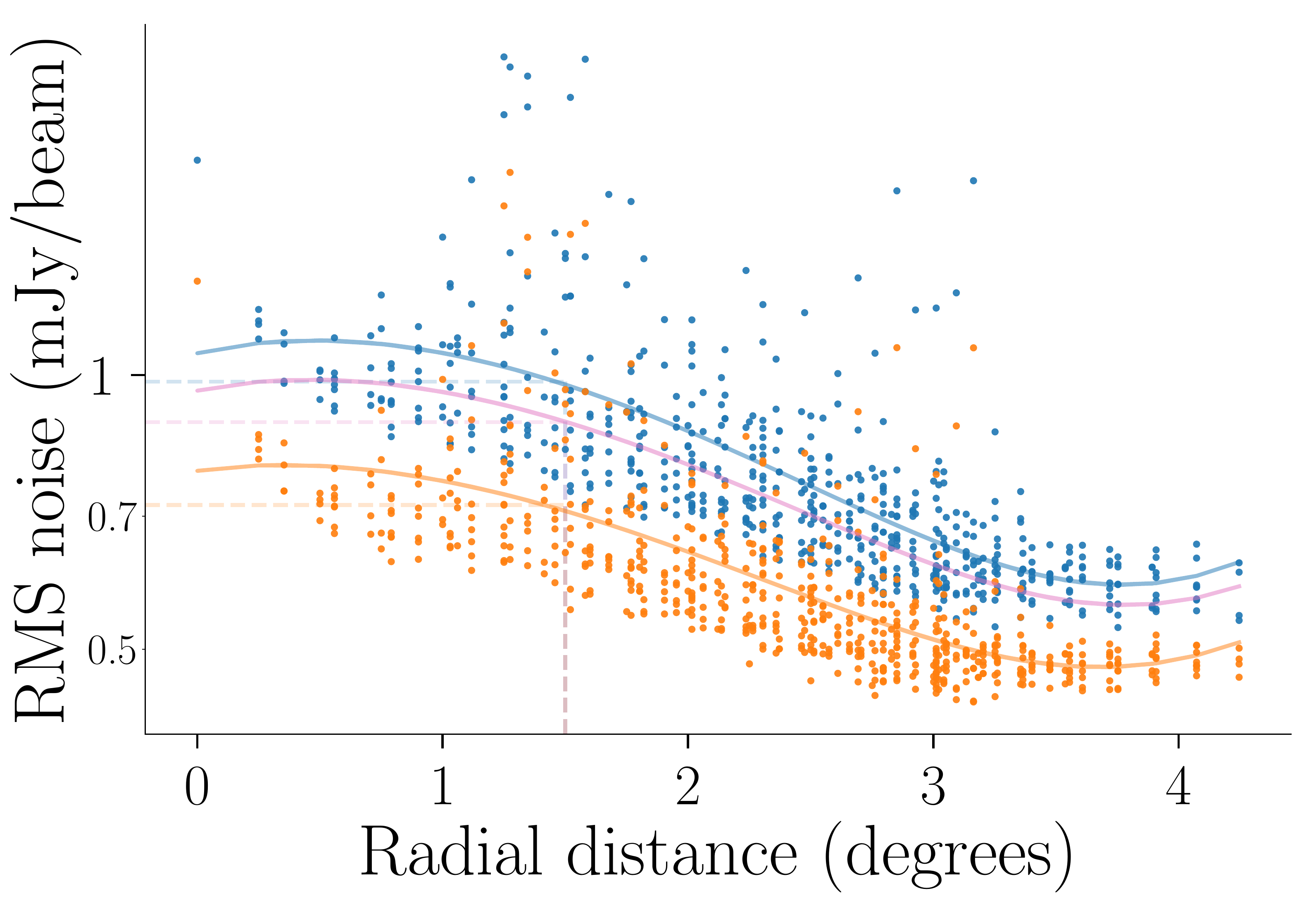}{0.20\linewidth}{(iv-b)}
   \fig{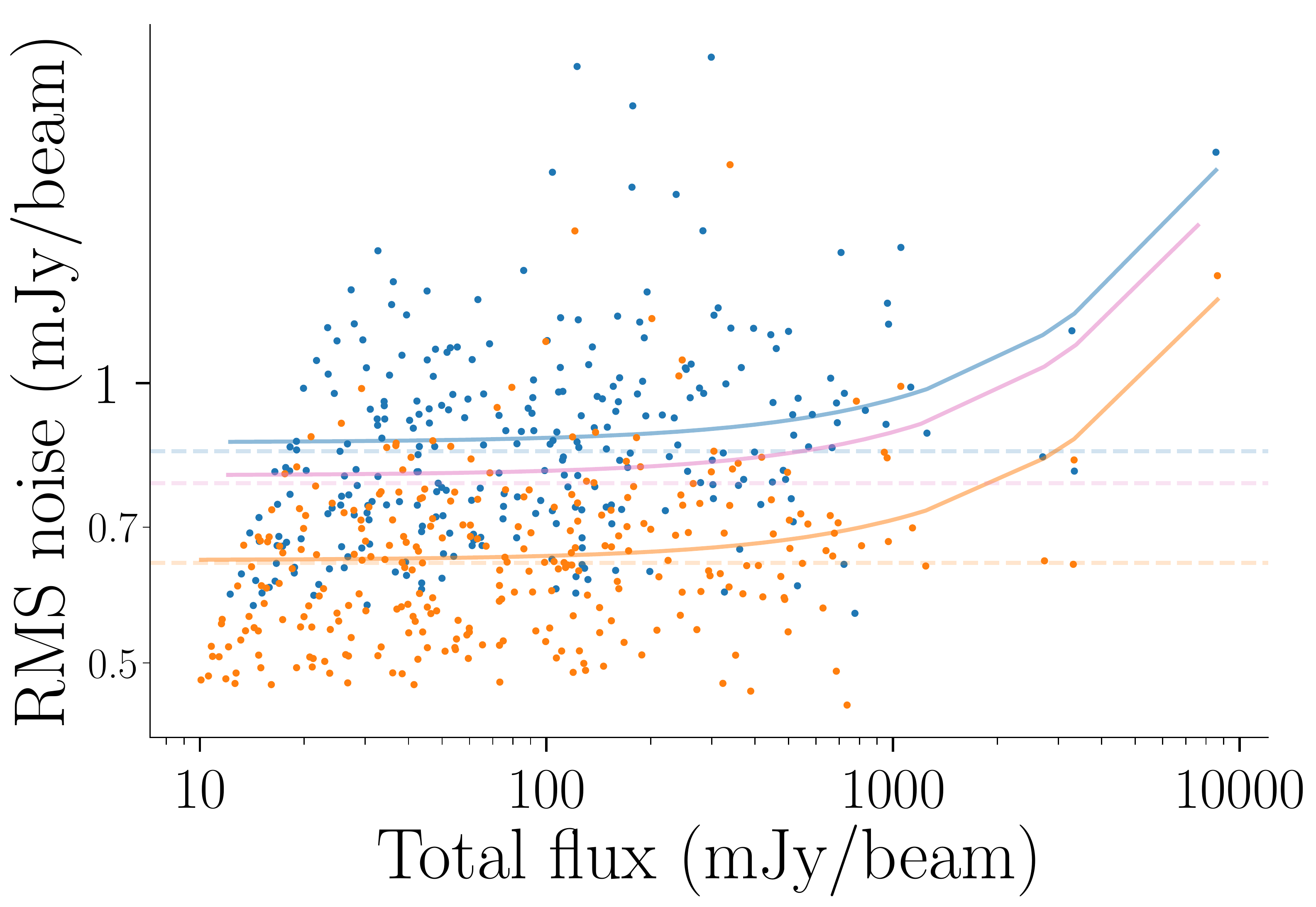}{0.20\linewidth}{(iv-c)}
   \fig{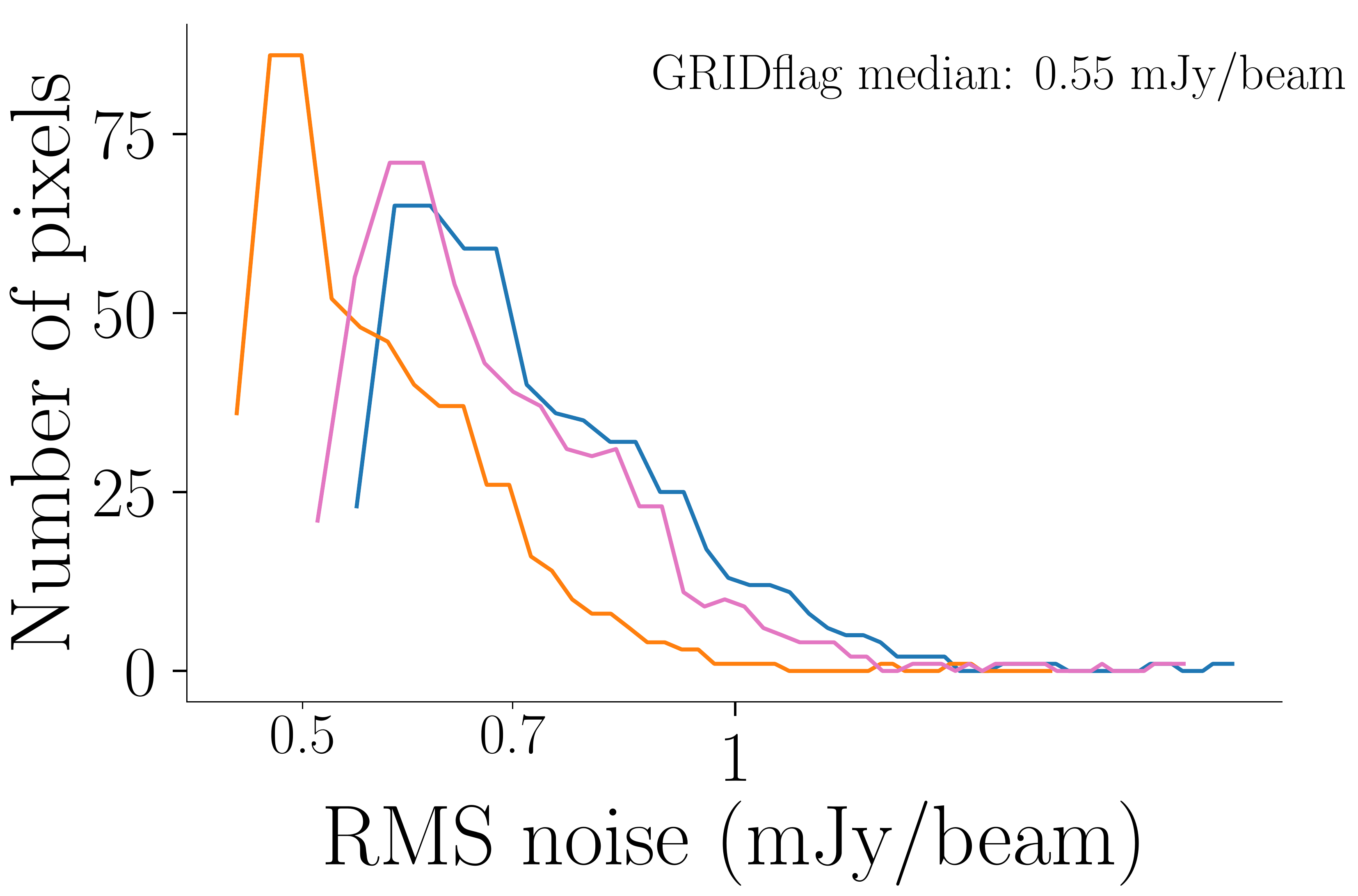}{0.20\linewidth}{(iv-d)}
   \fig{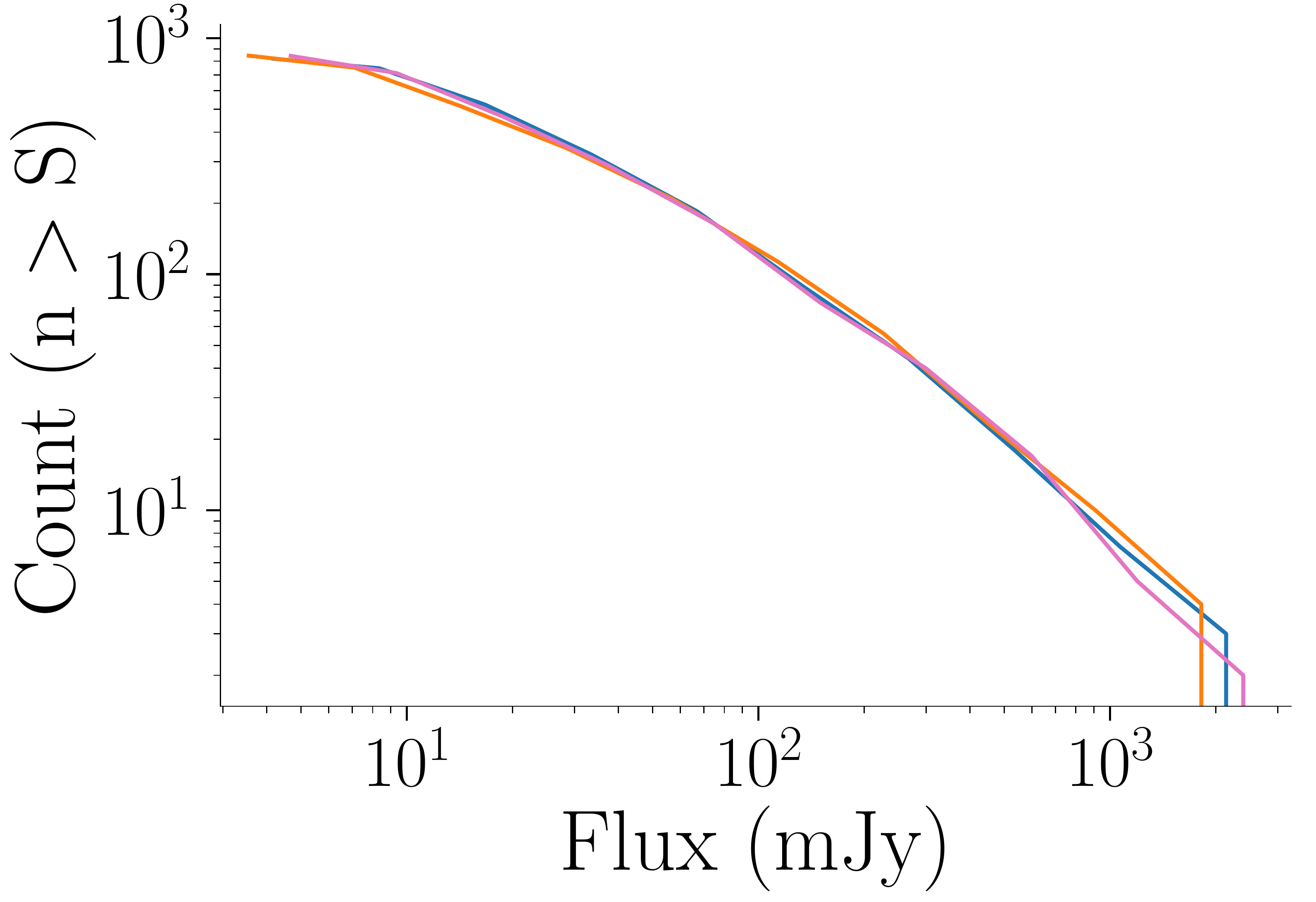}{0.20\linewidth}{(iv-e)}
}
   \gridline{
      \text{A2163}
   }
   \gridline{
      \fig{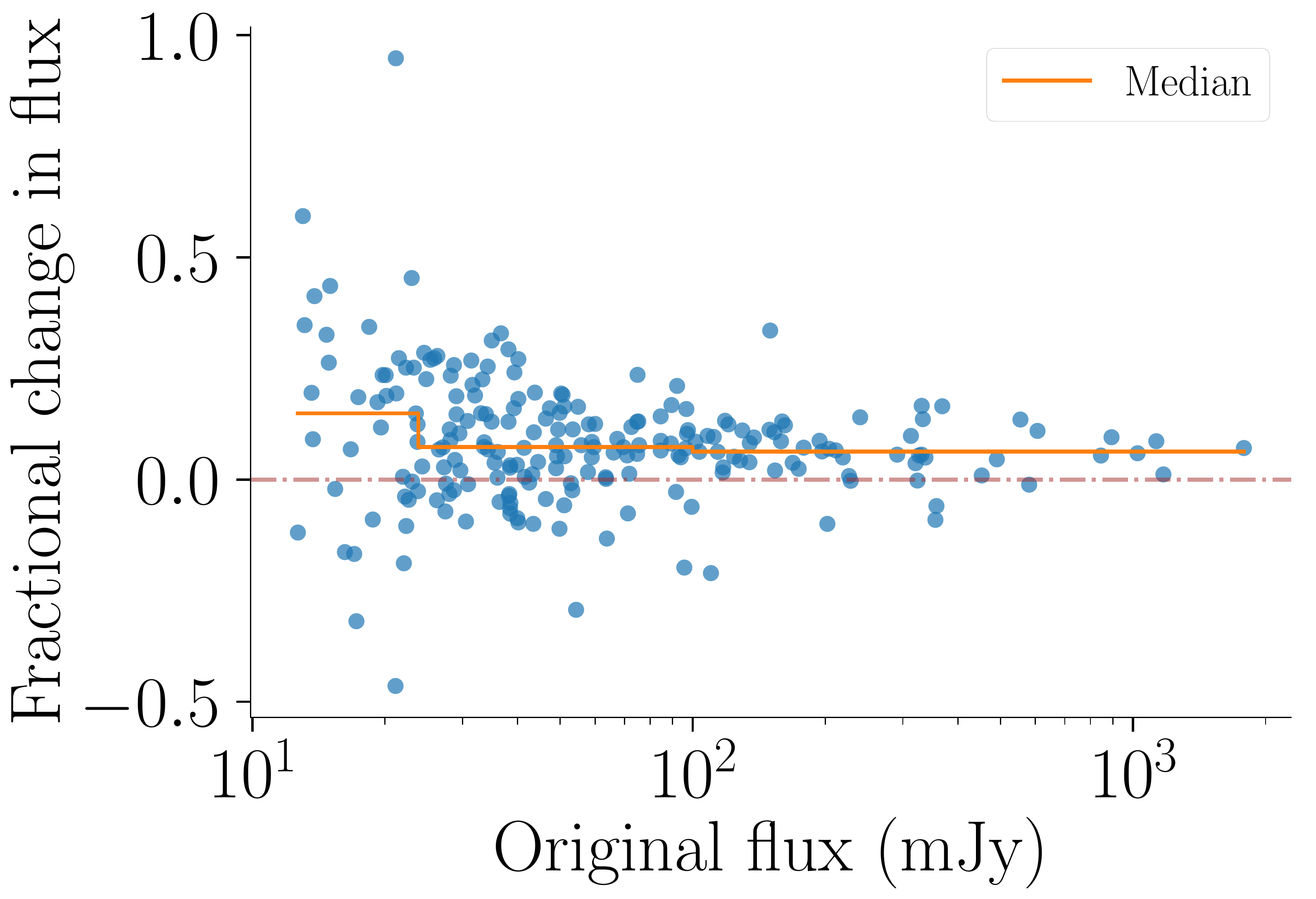}{0.20\linewidth}{(v-a)}
   \fig{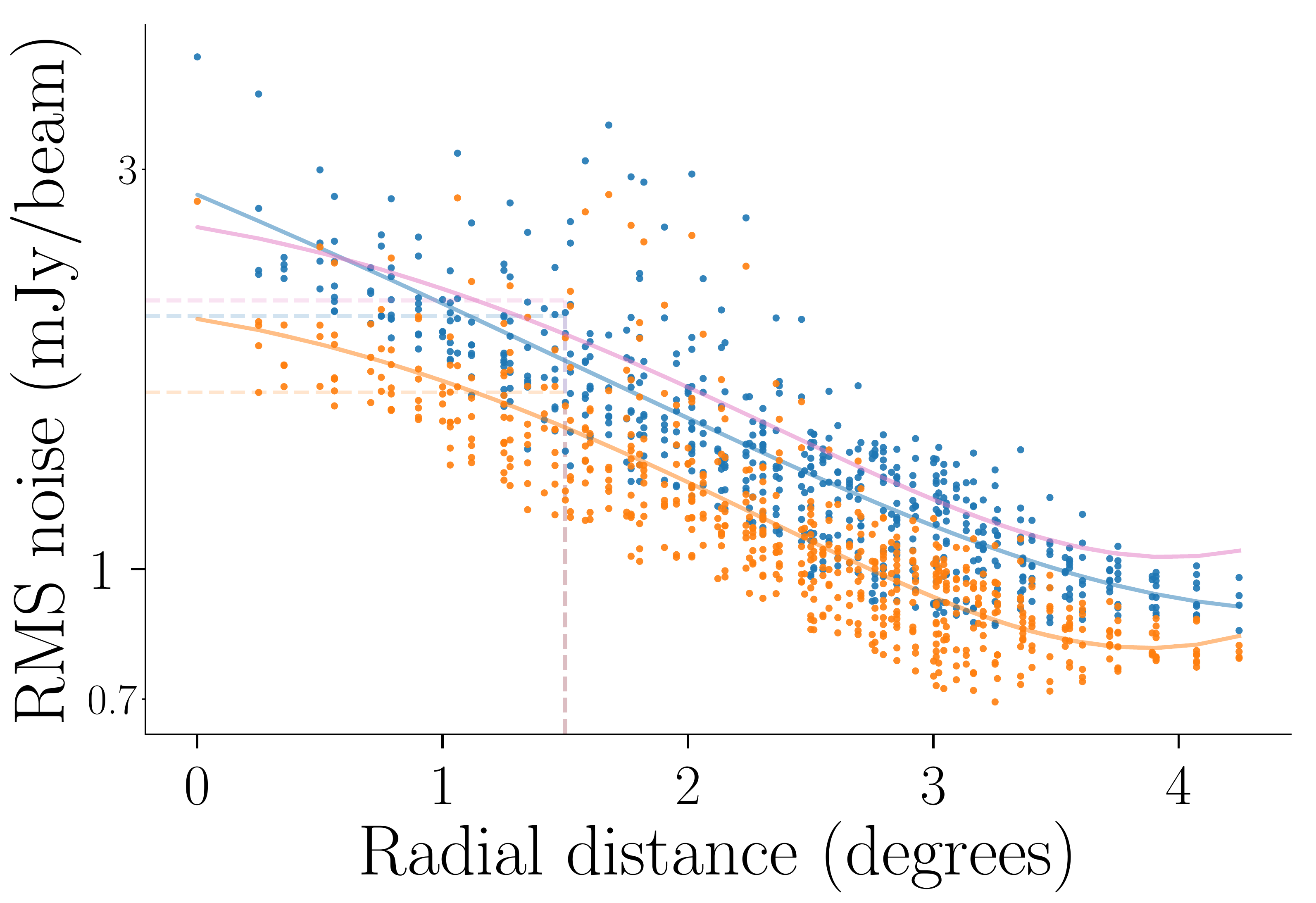}{0.20\linewidth}{(v-b)}
   \fig{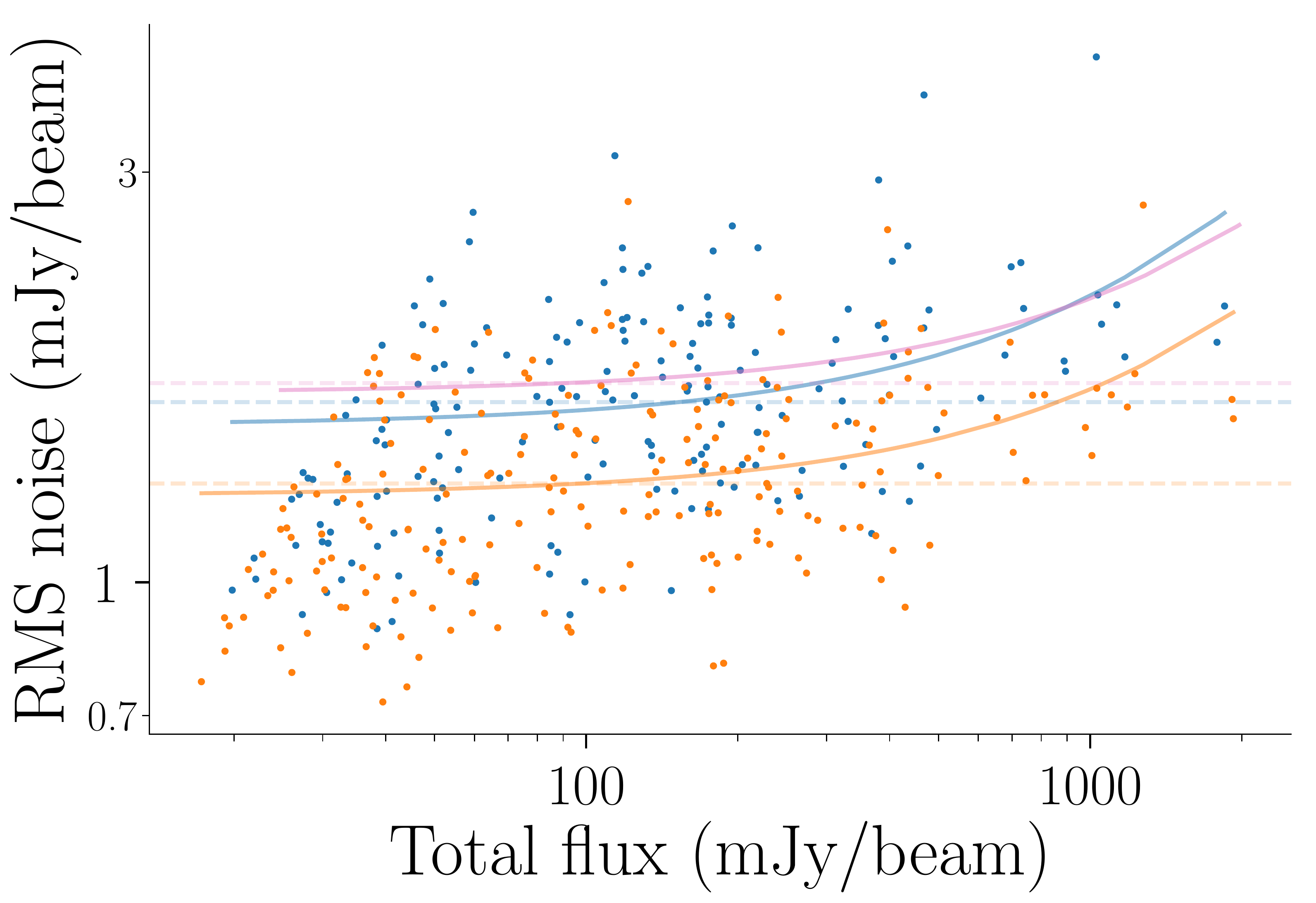}{0.20\linewidth}{(v-c)}
   \fig{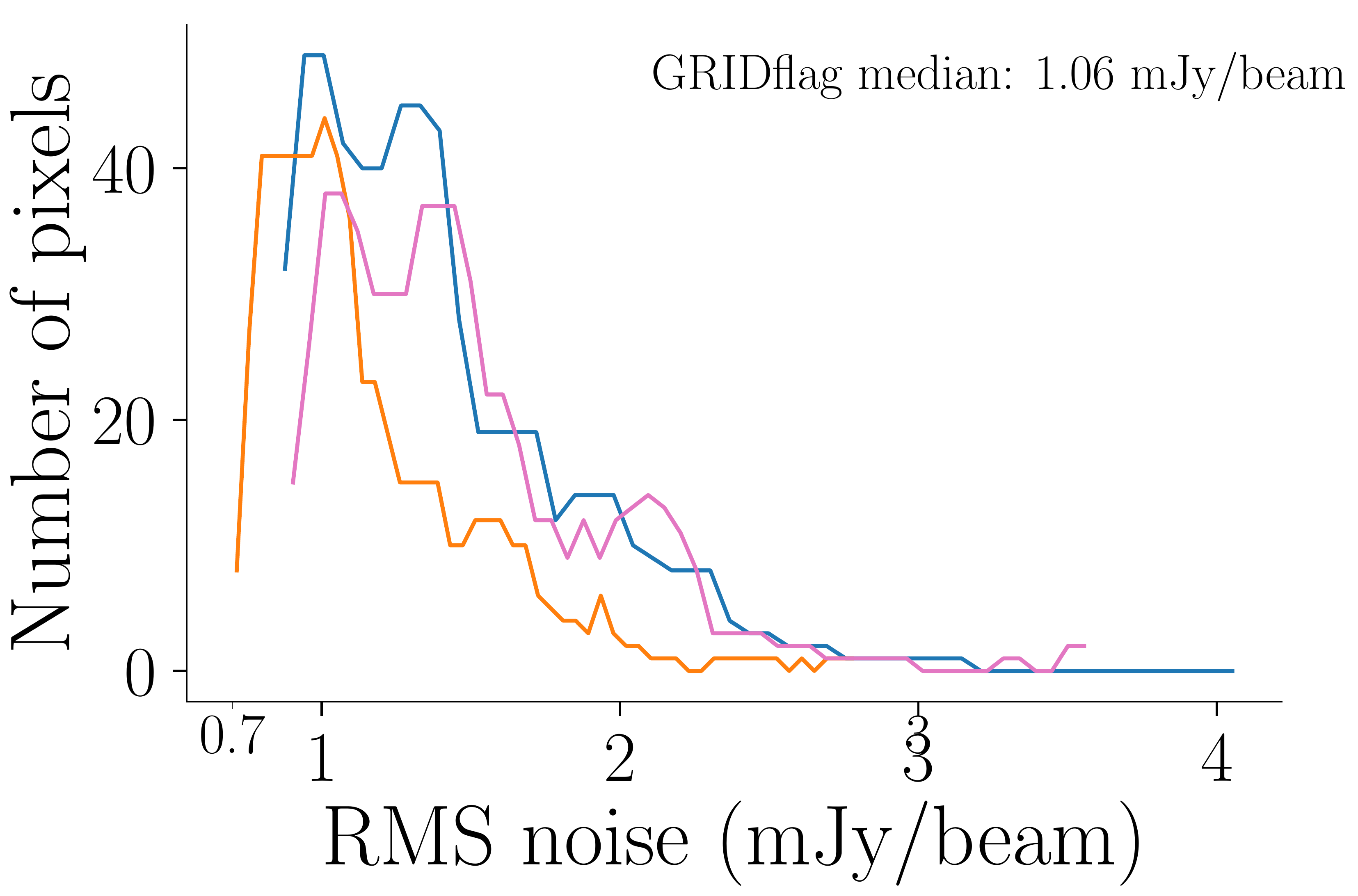}{0.20\linewidth}{(v-d)}
   \fig{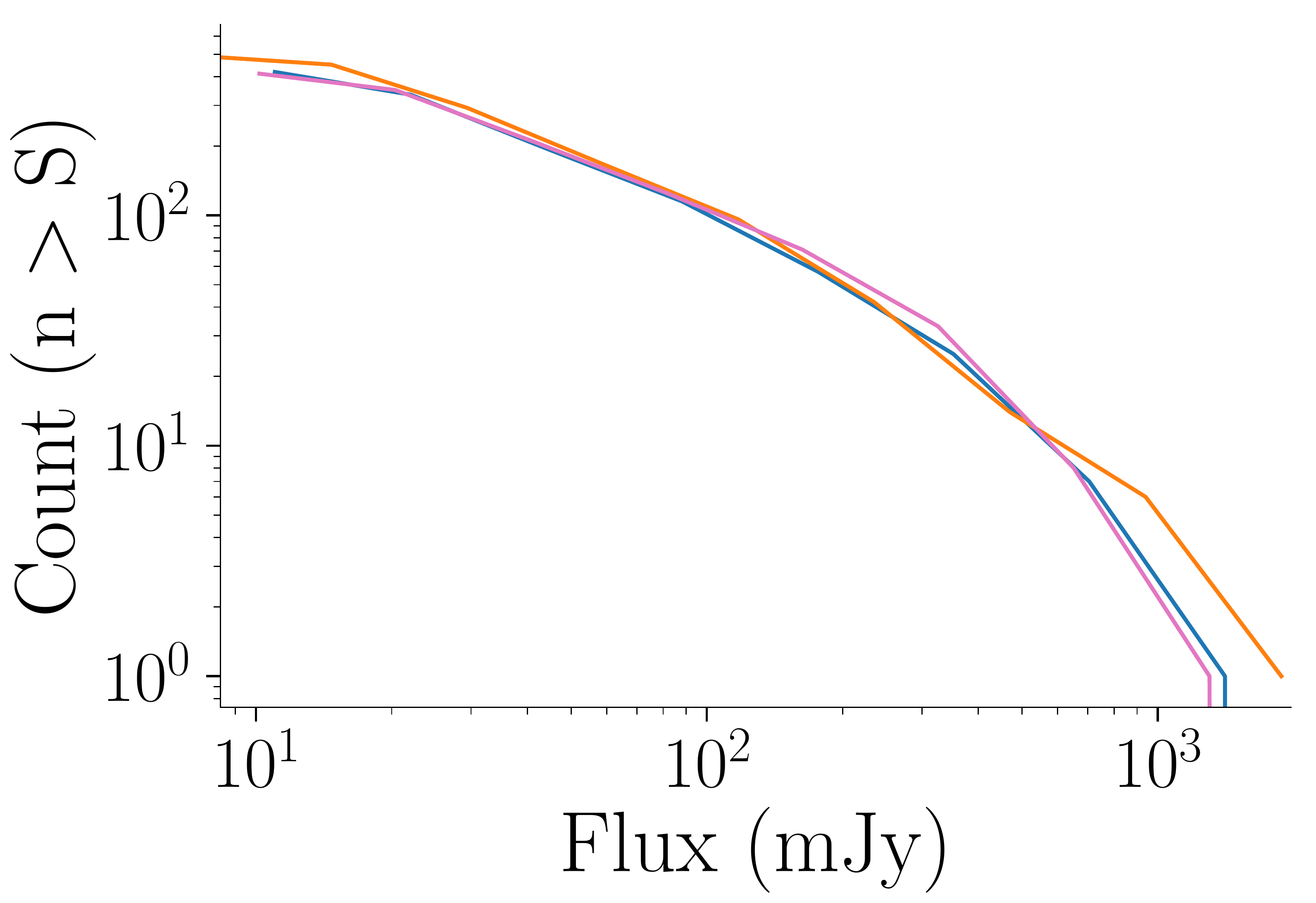}{0.20\linewidth}{(v-e)}
}

   \caption{Comparison of images generated using IPFLAG, AOflagger and with
   neither. Each row corresponds to a different target, while the 5 columns show
   from left to right, (a) Fractional change in flux as a function of
   original (pre-flagging) flux - only for IPFLAG; the solid line shows the
   median change separately for the strong (S $>$ 100 mJy), intermediate
   ($25\sigma < S < 100$ mJy) and faint (S $< 25\sigma$) sources. (b) Image
   noise as a function of radial distance from the phase centre (c) Image noise
   as a function of neighbourhood source flux (d) Histogram of image noise
   values from across the field and (e) Integral source counts. The results for
   IPFLAG, AOFlagger and ``Neither'' are shown in orange, pink and blue,
   respectively. Application of IPFLAG resulted in the detection of fainter
   sources as evidenced by the extension of source counts to lower flux
   densities.}
   \label{fig:all_plots}
\end{figure*}

\paragraph{Point source fluxes} The plots in Figure \ref{fig:all_plots}a
show the fractional change in flux density (FCF) as a function of the original
(un-flagged) flux density. These flux densities were compared before primary beam
correction. We analysed the FCF in three flux density regimes, \emph{viz.} S $>
100$ mJy, $25\sigma < \text{S} < 100$ mJy and S $< 25\sigma$.  $25\sigma$ is the
flux density above which source counts are typically $\sim$90\% complete (e.g.,
\citealt[][]{intema2017gmrt}) and as such defines the limit of the reliability
of source catalogs for statistical studies. The FCF ranges from -1\% to +6\% for
the strongest sources. The increase in flux may be expected due to an
improvement in calibration after the RFI has been flagged.  The mean FCF for
intermediate sources is smaller than that for the strong sources by -6\% to
+1\%. This trend in reduction of flux density continues to the faintest sources
detected. We saw this reduction with all the others flaggers as well (AOFlagger,
RFlag, and TFCrop), and both the source detection algorithms (PyBDSF and aegean;
\citealt[][]{mohan_pybdsf_2015}, \citealt[][]{hancock2012compact}) used.
Anticipating that this discrepancy may be a result of a systematic shift in the
background level we analysed the residuals in the immediate vicinity of the
detected sources before and after flagging, but did not detect any such
systematic offsets.  In summary, this effect is independent of the RFI flagging
algorithm used; it does not seem to affect the flux density scale since the
brightest sources are not affected; it must have something to do with the CLEAN
algorithm that we use, which comes with CASA; the change in flux density of the
intermediate sources is much smaller than the reduction in image noise. At the
moment we have no explanation for this effect but it may be something similar to
the CLEAN bias whose impact is most obvious for the faintest sources
\citep{condon1998nrao, cohen2007vla}. We recall that a similar effect was
observed while comparing our flux densities with that of TGSS-ADR in Figure
\ref{fig:flux_scale}.

\noindent
\paragraph{Image noise as a function of radial distance}

In general the image noise is known to reduce with radial distance.  A 200x200
pixel box was used to calculate the robust RMS at 625 locations across the
image. The plots in column b of Figure \ref{fig:all_plots} show both the
scatter and the trendline for the unflagged and IPFLAG data;
the AOFlagger data is only represented by the trendline in the interest of
clarity. The plots show that IPFLAG has reduced the noise all across
the image.

\noindent
\paragraph{Image noise as a function of source flux}
Column c of Figure \ref{fig:all_plots} shows a plot of the image noise
plotted against the cumulative flux density within each of the 625 boxes
mentioned earlier. Normally, a higher level of local artefacts is
expected in the presence of strong point sources.

\noindent
\paragraph{Noise Histogram}

Column d of Figure \ref{fig:all_plots} shows the histogram of the noise
across the field. There is a clear shift in the distribution of RMS noise
to lower values.

\noindent
\paragraph{Source Counts}
The ultimate metric for image improvement is an increase in
source detection at lower flux levels. Figure \ref{fig:all_plots}e shows
that IPFLAG passes this criterion by detecting faint sources to the
expected depth. The plot shows the cumulative histogram of the number
of sources detected as a function of flux density \emph{i.e.,} N($>$ S) which
shows clearly that the excess detections come from lower flux densities.
{Table \ref{table:image_parameters} provides the number of sources
detected for each field. In all cases, IPFLAG has resulted in the detection
of more sources.

\noindent
\paragraph{Artefacts} Figure \ref{fig:postage_stamps} shows the
reduction in artefacts in the field after the application of IPFLAG.

\begin{figure*}[htbp]
   \gridline{
      \fig{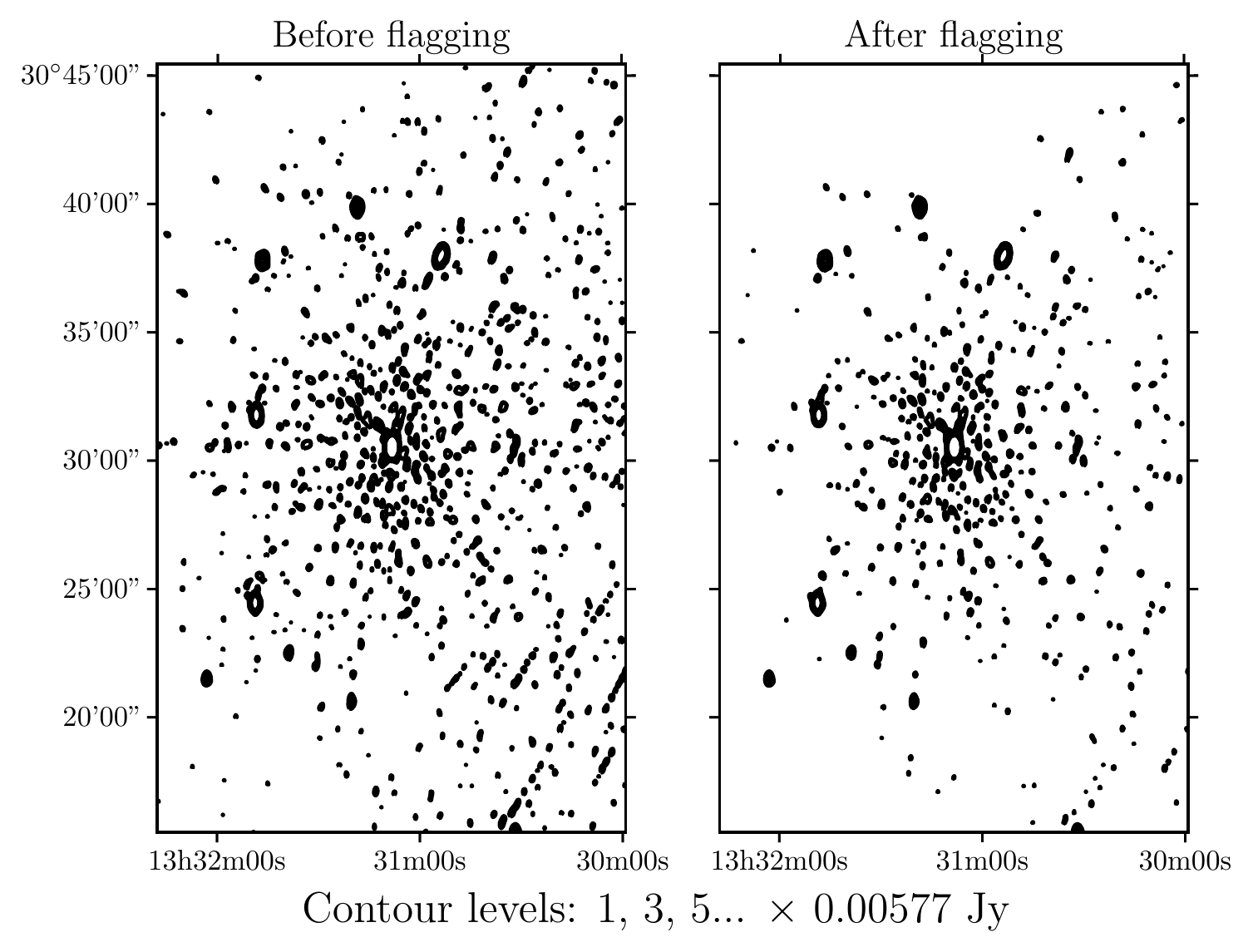}{0.3\linewidth}{3C286}
   \fig{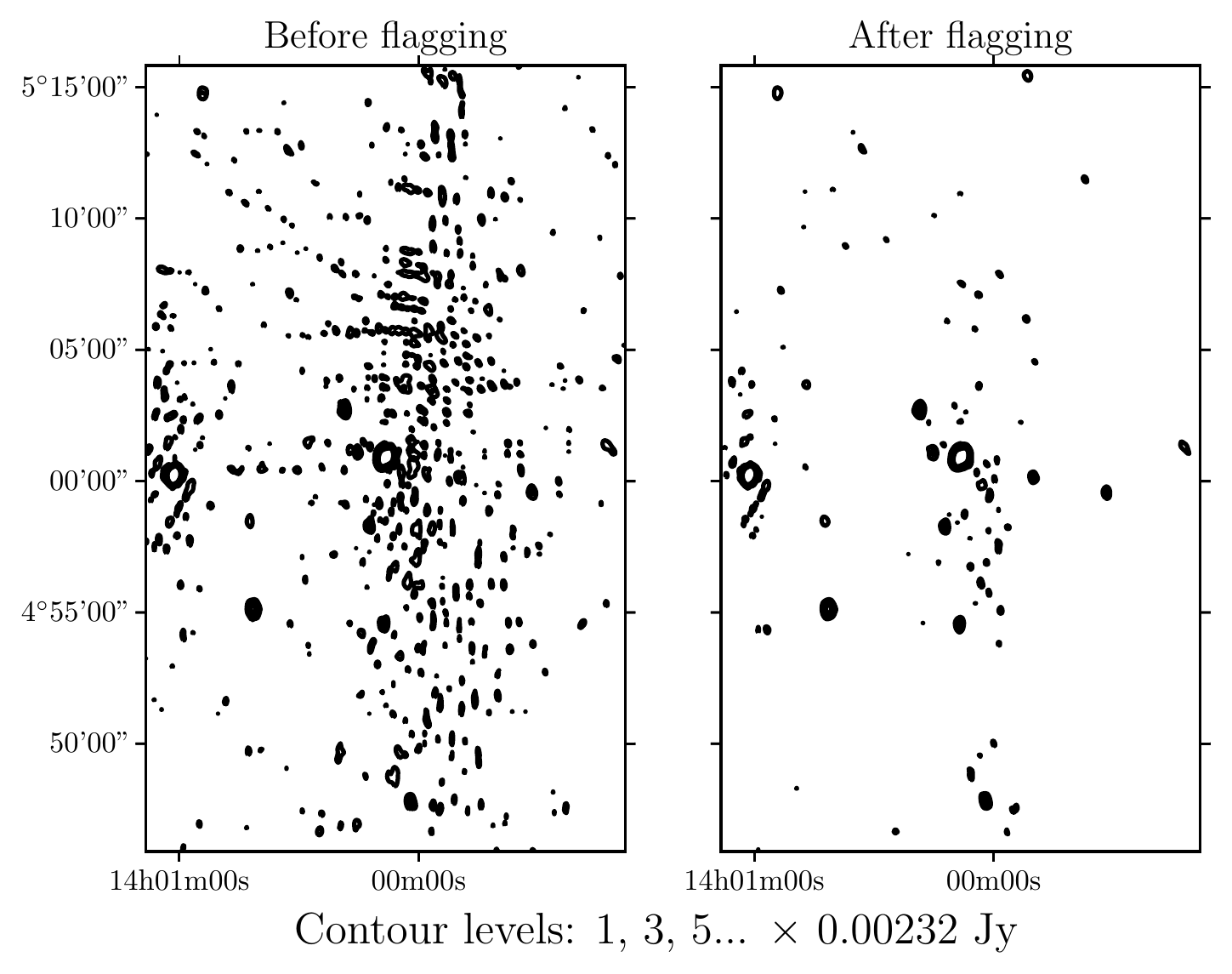}{0.3\linewidth}{VIRMOSC}
   \fig{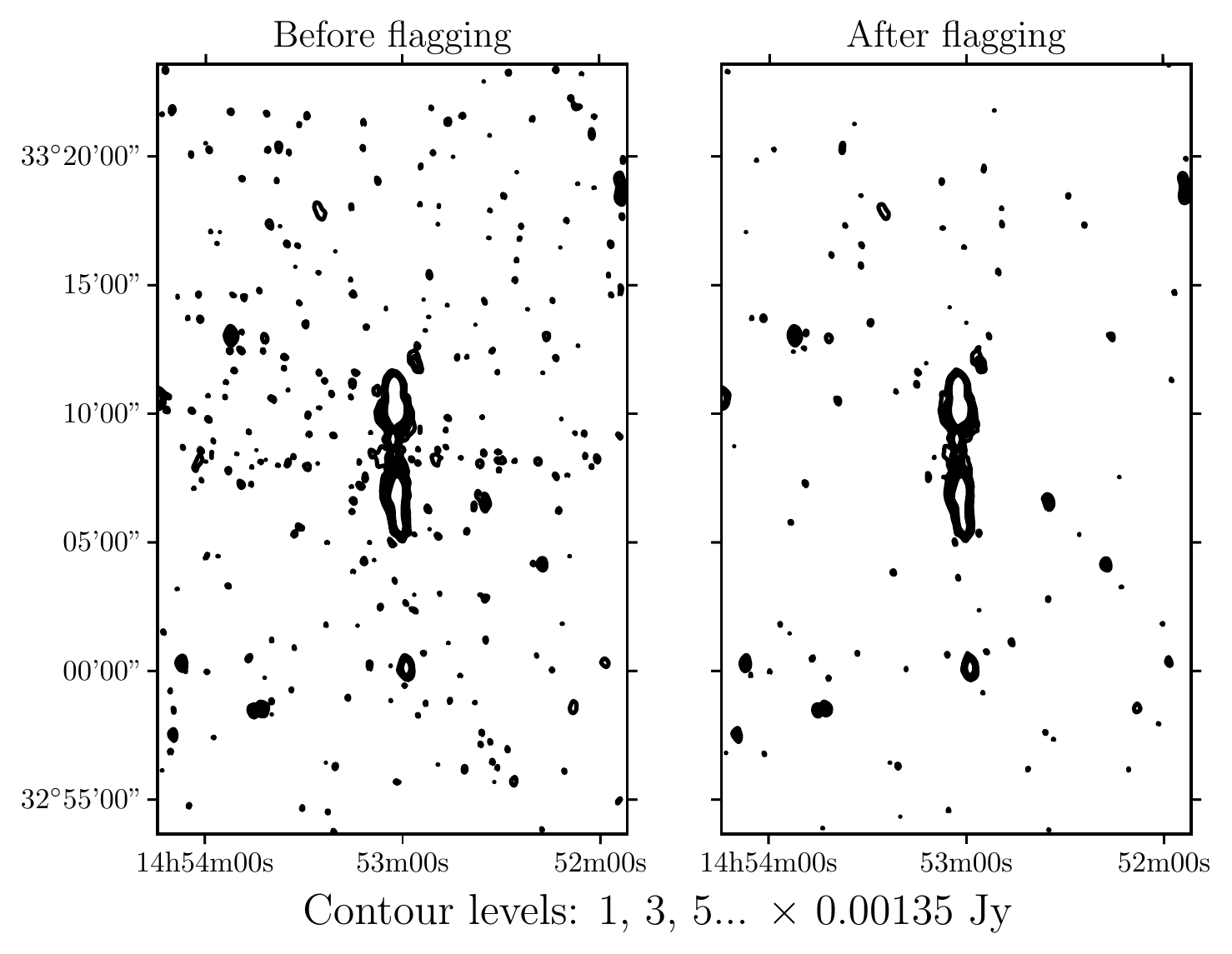}{0.3\linewidth}{J1453+3308}
}
   \gridline{
      \fig{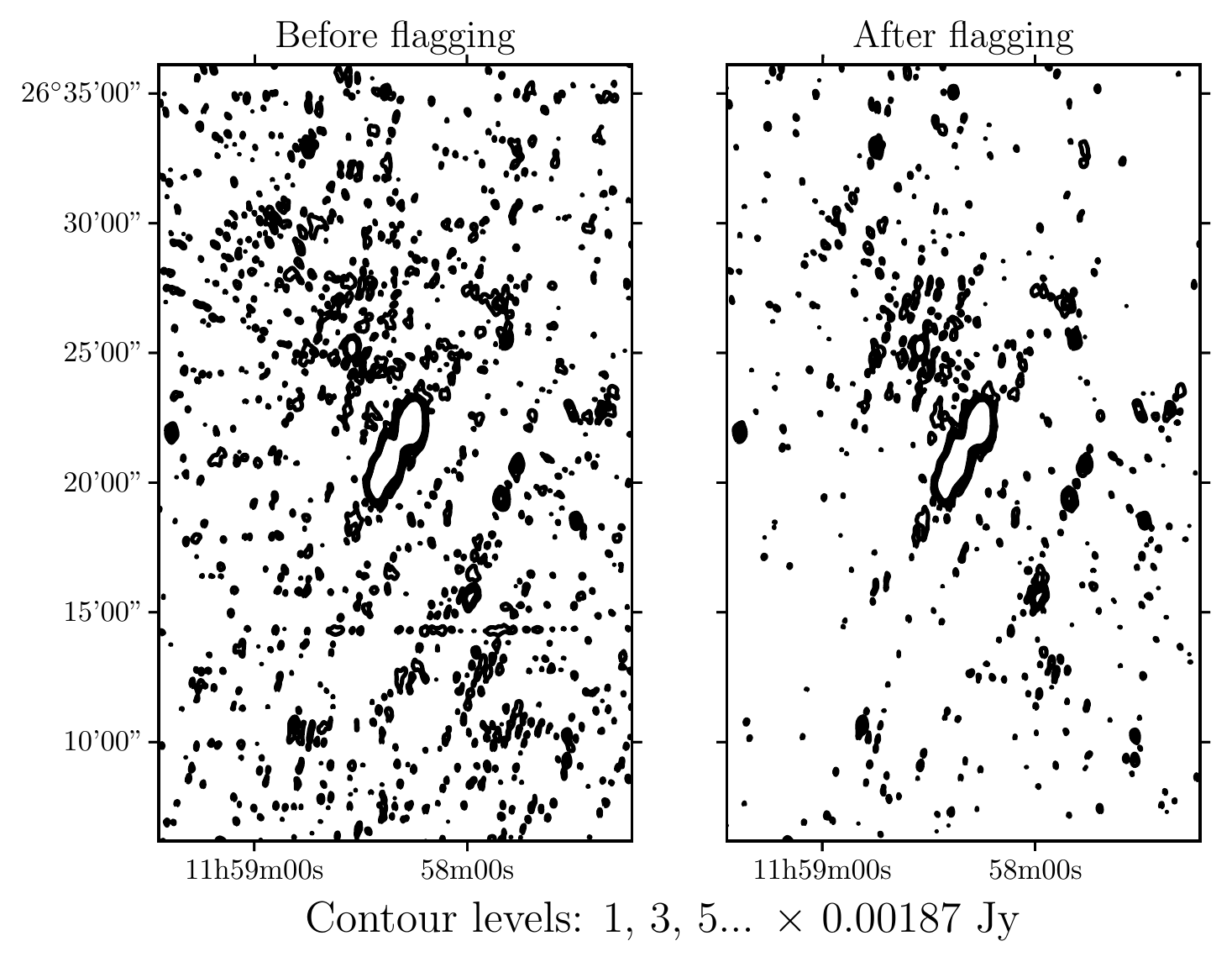}{0.3\linewidth}{J1158+2621}
   \fig{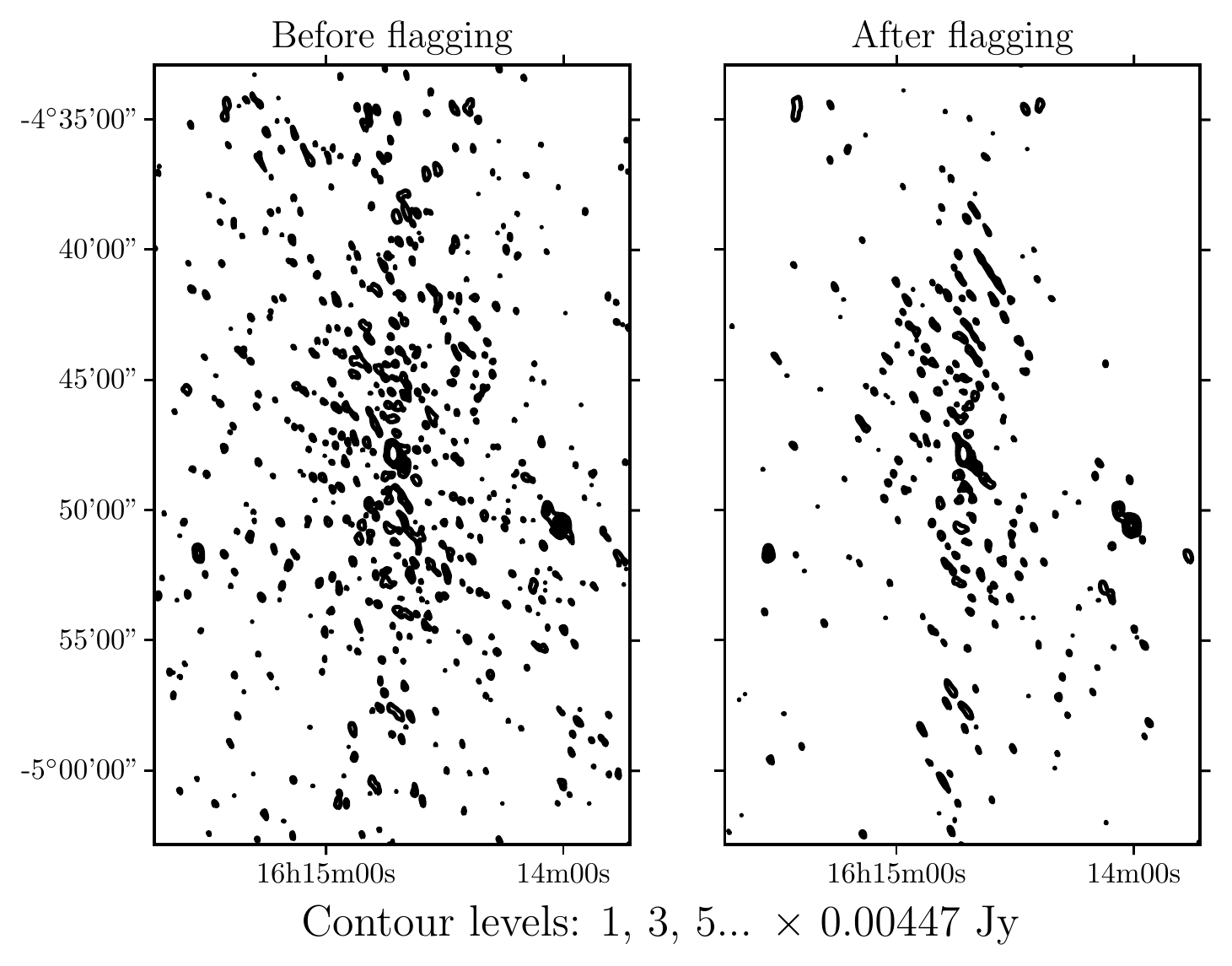}{0.3\linewidth}{A2163}
}

\caption{A comparison of artifacts in the inner parts of images (1800x2700
arcsec$^2$, pixel size 4.5$\arcsec$) - except for the A2163 field where an
off-centre source is shown -  made with and without the application of
IPFLAG. The contours levels are indicated below each plot. The
unit contour level corresponds to three times the local standard deviation
of the IPFLAG image in each case. All fields show a substantial
reduction in artefacts on application of IPFLAG.}

   \label{fig:postage_stamps}
\end{figure*}

\section{Conclusions}

The IPFLAG procedures have worked well across a variety of datasets
substantially reducing the image noise and detecting fainter sources.  While we
tested the algorithms on GMRT data they should be as effective for any
interferometric array in which the UV plane contains a large number of
visibility samples.

We are unaware of any other algorithm which approaches RFI flagging in
the manner that GRIDflag does. TCflag is similar to several other algorithms
like AOflagger, TFcrop, Rflag, etc, which flag localised RFI in the
time-channel plane of an individual baseline while differing in the manner of
estimating the background. The combined application of GRIDflag+TCflag, which
are components of our RFI pipeline IPFLAG outperforms the other algorithms.
AOFlagger was the closest in terms of performance and we recognise that a more
experienced user of AOflagger may be able to obtain better results by
optimising its many tuneable parameters for GMRT data.

On the other hand IPFLAG has only six tuneable parameters in all.  These values
did not require much optimisation and the same values were used for all the
fields tested. In fact, of the six, the UV-bin size selects itself from the
field of view and the three noise thresholds were set to ``universal'' default
values (at 3$\sigma$). The width of the UV-annuli was set by simply distributing
the visibilities approximately equally across the 3 annuli used.  While the
values may require some changes for other frequencies and interferometers, we
believe that they will not need to be tuned for different fields.

On application of IPFLAG our images have consistently reached an RMS noise $<$1
mJy/beam, with a a corresponding increase in source detection; typical GMRT 150
MHz images have hitherto reached an RMS noise of 1.5 - 5 mJy/beam, with
exceptional effort yielding 0.7 mJy/beam \citep[e.g.][]{ishwara2010deep}. In 3
of the five images we have reached an image noise of 0.38-0.55 mJy/beam
which is only a factor of $\sim$2 above the theoretical confusion limit. Even
in the case of 3C286, an exceptionally bright flux density calibrator, we have
reached 1.0 mJy/beam and the peak to RMS noise ratio is in excess of 21,000
which is unprecedented for GMRT 150 MHz images.

The fact that GRIDflag reduces the loss of UV coverage in the gridded UV plane
means that the reduction in RFI is not offset by a corresponding increase in
the sidelobes of the synthesised beam. The efficacy of the
algorithm depends on the level of redundancy in the gridded visibility plane.
The GMRT has good coverage of the shorter spacings, particularly at low
frequencies \citep{swarup_giant_1991} and this coverage is due to get better
with the upgraded GMRT \citep{gupta2017upgraded}.

The algorithms also worked at higher frequencies (325 MHz and 610 MHz) but
the improvement was not as substantial. We think that this is because of the
weaker RFI environment at these frequencies. However, even these weak RFI
environments could substantially impact the performance of ultra-deep
imaging projects like the MIGHTEE \citep{jarvis2017meerkat}. Further, we
believe that GRIDflag is particularly tailored towards multi-epoch
observations such as the MIGHTEE wherein the same UV grids are sampled day
after day.

These algorithms will not work in the presence of persistent, broadband RFI.
However, other algorithms are available for these situations
\citep[e.g.][used in this paper]{athreya_new_2009}. We think that the issues
which remain to be addressed are non-isoplanatic ionosphere and an asymmetric
antenna primary beam.  Applying our algorithms in conjunction with  schemes
like SPAM \citep{intema_ionospheric_2009} should yield further improvement of
image quality.

\acknowledgements
We thank the staff of the GMRT who have made these observations possible.
GMRT is run by the National Centre for Radio Astrophysics of the Tata
Institute of Fundamental Research. Discussions with Drs. Sanjay Bhatnagar and
Ishwara-Chandra contributed to this work. We thank the referee for many
suggestions which have considerably improved the manuscript.

\facility{GMRT}

\bibliography{references}

\end{document}